\documentclass[final,5p,times,twocolumn]{elsarticle}
\let\today\relax
\makeatletter
\def\ps@pprintTitle{%
    \let\@oddhead\@empty
    \let\@evenhead\@empty
    \def\@oddfoot{\footnotesize\itshape
         {Preprint to be published in NeuroImage} \hfill\today}%
    \let\@evenfoot\@oddfoot
    }
\makeatother
\usepackage{graphicx}
\usepackage{amssymb}
\usepackage{lipsum}
\usepackage{mathtools}
\usepackage{makecell}
\usepackage{subfigure}
\usepackage{amsmath}
\DeclareMathOperator*{\argmin}{argmin}

\usepackage{natbib}
\usepackage{ifpdf}
\ifpdf
 \usepackage[pdftex]{hyperref}
\else
 \newcommand{\href}[2]{#2}
\fi
\biboptions{semicolon,round}

\setcitestyle{authoryear,open={(},close={)}}

\hypersetup{
     colorlinks   = true,
     citecolor    = blue
}

\journal{NeuroImage}

\begin{document}

\begin{frontmatter}


\title{Neuro4Neuro: A neural network approach for neural tract segmentation using large-scale population-based diffusion imaging}

\author[1,2]{Bo Li}
\author[2,4]{Marius de Groot}
\author[2]{Rebecca M. E. Steketee}
\author[2,7]{Rozanna Meijboom}
\author[2]{Marion Smits}
\author[2,4]{Meike W. Vernooij}
\author[2,4,5]{M. Arfan Ikram}
\author[1]{Jiren Liu}
\author[2,6]{Wiro J. Niessen}
\author[2]{Esther E. Bron}
\date{}
\address[1]{Sino-Dutch Biomedical and Information Engineering School, Northeastern University, Shenyang, China}
\address[2]{Department of Radiology and Nuclear Medicine, Erasmus MC, Rotterdam, the Netherlands}
\address[4]{Department of Epidemiology, Erasmus MC, Rotterdam, the Netherlands}
\address[5]{Department of Neurology, Erasmus MC, Rotterdam, the Netherlands}
\address[6]{Imaging Physics, Applied Sciences, Delft University of Technology, the Netherlands}
\address[7]{Centre for Clinical Brain Sciences, University of Edinburgh, United Kingdom}

\begin{abstract}

Subtle changes in white matter (WM) microstructure have been associated with normal aging and neurodegeneration. To study these associations in more detail, it is highly important that the WM tracts can be accurately and reproducibly characterized from brain diffusion MRI. In addition, to enable analysis of WM tracts in large datasets and in clinical practice it is essential to have methodology that is fast and easy to apply. This work therefore presents a new approach for WM tract segmentation: Neuro4Neuro, that is capable of direct extraction of WM tracts from diffusion tensor images using convolutional neural network (CNN). This 3D end-to-end method is trained to segment 25 WM tracts in aging individuals from a large population-based study (N=9752, 1.5T MRI). The proposed method showed good segmentation performance and high reproducibility, i.e., a high spatial agreement (Cohen's kappa, $\kappa=0.72-0.83$) and a low scan-rescan error in tract-specific diffusion measures (e.g., fractional anisotropy: $\epsilon=1\%-5\%$). The reproducibility of the proposed method was higher than that of a tractography-based segmentation algorithm, while being orders of magnitude faster (0.5s to segment one tract).  In addition, we showed that the method successfully generalizes to diffusion scans from an external dementia dataset (N=58, 3T MRI). In two proof-of-principle experiments, we associated WM microstructure obtained using the proposed method with age in a normal elderly population, and with disease subtypes in a dementia cohort. In concordance with the literature, results showed a widespread reduction of microstructural organization with aging and substantial group-wise microstructure differences between dementia subtypes. In conclusion, we presented a highly reproducible and fast method for WM tract segmentation that has the potential of being used in large-scale studies and clinical practice. 
\end{abstract}



\end{frontmatter}



\newcommand\blfootnote[1]{%
  \begingroup
  \renewcommand\thefootnote{}\footnote{#1}%
  \addtocounter{footnote}{-1}%
  \endgroup
}
\blfootnote{\textit{Abbreviations}: MRI, Magnetic Resonance Imaging; DTI, Diffusion Tensor Imaging; FA, Fractional Anisotropy; ICV, Intracranial Volume; MD, Mean Diffusivity; ROIs, Regions of Interest; SD, Standard Deviation; TE, Echo Time; TR, Repetition Time}

\DeclareRobustCommand{\de}[3]{#3}

\newcommand{\beginsupplement}{%
        \setcounter{table}{0}
        \renewcommand{\thetable}{S\arabic{table}}%
        \setcounter{figure}{0}
        \renewcommand{\thefigure}{S\arabic{figure}}%
     }

\section{Introduction}
\label{S:1}

Changes in the micro- and macrostructure of brain white matter (WM) are known to be related to cognitive impairment and neurodegeneration \citep{fellgiebel2005color,abe2002normal,vernooij2008white}. The WM consists of axonal fibers that enable communication between brain regions and can be functionally grouped into WM tracts. To improve the understanding of WM tracts and their involvement in the processes of neurodegeneration in aging and disease, it is essential to segment them and quantify their microstructure with high accuracy and reproducibility. This is however non-trivial because WM tracts cannot be identified directly from diffusion magnetic resonance imaging (dMRI) and because their anatomy can be complex.

Most WM tract segmentation methods are based on reconstruction of potential WM fibers by tractography on dMRI. Those tractography-based segmentation methods can be grouped into three categories: semi-automatic, atlas-based and clustering methods \citep{sydnor2018comparison}. Semi-automatic methods use automated tractography assisted by manual delineations of regions-of-interest (ROIs) \citep{mori2005mri}. This however requires substantial neuroanatomical knowledge, is time consuming and is highly operator-dependent. Especially in tracts with complex geometry, brain regions with crossing fibers and data with low quality, semi-automatic methods have shown limited reproducibility \citep{wakana2007reproducibility}. As the name implies, atlas-based segmentation methods use anatomical priors propagated from single or multiple atlases for tractography initialization and/or pruning \citep{wakana2004fiber,lawes2008atlas,hua2008tract,suarez2012automated,de2015tract,wassermann2016white,yendiki2016joint,yendiki2011automated,zollei2019tracts}. Clustering methods are fully automatic as well, in which tractography streamlines are grouped into tracts based on combined metrics of geometric trajectories, distance similarity, homology across hemispheres, consistency across subjects, or additional anatomical constraints like shape priors and spatial priors \citep{o2007automatic, prasad2014automatic,jin2014automatic,garyfallidis2017recognition}.

Another class of WM tract segmentation methods are of machine learning strategies using fiber-based classification \citep{poulin2017learn,lam2018trafic,gupta2018fibernet,jha2019fs2net,zhang2019deep,liu2019deepbundle} or voxel-wise classification \citep{bazin2011direct,ratnarajah2014multi,wasserthal2018tractseg,li2018reproducible,li2019hybrid}. Unlike the previously described approaches, voxel-wise classification methods do not rely on tractography, but directly label voxels as specific tracts based on their diffusion information. Recently, deep-learning techniques, in particular convolutional neural networks (CNN), have emerged as a powerful tool and shown to be very successful. CNN-based methods tackle segmentation tasks as the estimation of a parametric map-function between inputs and outputs, where the map function is modeled by a series of convolution and non-linearity operations. To estimate parameters - the weights of convolution kernels, CNN models are globally optimized over training datasets aiming at minimizing a loss function that measures difference from objectives. Given the advantage of segmentation accuracy and efficiency, CNN-based methods have been widely favored in image analysis field. For WM tract analysis, the effect of approach configurations, temporal consistency, and pre-clinical applicability have however barely been explored on large-scale imaging datasets.

In this paper, we developed and evaluated a 3D CNN method for WM tract segmentation: Neuro4Neuro. This method advances the state-of-the-art by being the first tract segmentation method that uses a 3D CNN. Furthermore, we utilize a large-scale dataset for optimizing the method and evaluating its potential of deep learning for WM tract segmentation. We quantitatively evaluated the method's accuracy and reproducibility, demonstrated its applicability for addressing  clinical research questions, and assessed its generalizability to an external patient dataset. This work is an extension of a previous conference article \citep{li2018reproducible}. In this extension, we improved preprocessing and the optimization experiments, extended validation from two to 25 tracts, and added a substantial number of evaluation experiments. The remainder of the paper is organized as follows: section \ref{S:2} presents the method including optimization experiments, section \ref{s:3} presents evaluation experiments and results, and section \ref{s:4} discusses the results and their implications.

\section{Neuro4Neuro}
\label{S:2}

\subsection{Materials and methods}
\label{ssec:2.1}

\subsubsection{Study population}
\label{sssec:2.1.1}
The Rotterdam Study is a prospective population-based study targeting causes and consequences of age-related diseases among 14,926 participants \citep{hofman2015rotterdam}. Since 2005, brain MRI has been incorporated in the core protocol. The Rotterdam study has been approved by the local medical ethics committee according to the Population Study Act Rotterdam Study, executed by the Ministry of Health, Welfare and Sports of the Netherlands. Written informed consent was obtained from all participants. 
In this work, we included 9752 dMRI scans from 5286 participants (age: $64.7 \pm 9.9$ years). For the optimization experiments (Section \ref{ssec:2.2}), a subset of 1082 scans ($D1$) was used, of which 864 scans were used for training ($D1_{train}$) and 218 scans for testing ($D1_{test}$). 

\subsubsection{MRI acquisition}
\label{sssec:2.1.2}
Scans were acquired on a 1.5T MRI scanner (GE Signa Excite). Diffusion weighted imaging (DWI) was acquired with following parameters: $TR/TE = 8575 ms/82.6 ms$; imaging matrix of $64\times96$ (zero-padded in k-space to $256\times256$) in a field of view (FOV) of $210\times210 mm^2$; 25 diffusion weighted volumes along non-collinear directions using a b-value of $1000 s/mm^2$ and three non-weighted volumes ($b=0 s/mm^2$). The voxel size was resampled from $3.3\times2.2\times3.5 mm^3$ to $1 mm^3$ during pre-processing. For T1-weighted images, the parameters were: $TR/TE = 13.8 ms/2.8 ms$; imaging matrix of $416\times 256$ in an FOV of $250\times250 mm^2$; The voxel size was $0.5\times0.5\times0.8 mm^3$.

\subsubsection{Image preprocessing}
\label{sssec:2.1.3}
DWI data were preprocessed using a previously described pipeline \citep{koppelmans2014global}. In short, motion and eddy currents were corrected by co-registering all diffusion weighted volumes to the averaged $b = 0$ volumes with Elastix \citep{klein2010elastix}. Diffusion tensors were estimated with a Levenberg-Marquard non-linear least-squares optimization algorithm, as available in ExploreDTI \citep{leemans2009exploredti}. We subsequently computed diffusion tensor imaging (DTI) measures: fractional anisotropy (FA), mean diffusivity (MD), axial diffusivity (L1), radial diffusivity (RD) and mode of anisotropy (MO). Due to noise, tensor estimation failed in a small proportion of voxels, resulting in significant outliers. Outlier voxels with a tensor norm (Frobenius norm) larger than $0.1 mm^2/s$ \citep{zhang2007high} were set to zero. The tensor images used as the input for proposed method were estimated and used in subject native diffusion space. The native diffusion space had a similar brain orientation for all subjects. No co-alignment with a standard orientation was performed. For each tract, an ROI was defined by taking the maximum bounding box based on the reference segmentation (Section \ref{sssec:2.1.5}). The magnitude of the tensors was scan-wise normalized to zero mean and a unit standard deviation. A brain tissue mask was obtained by combining WM and gray matter segmentations \citep{vrooman2007multi}.   

\subsubsection{Reference method}
\label{sssec:2.1.5}
For model training and evaluation, we generated the reference WM tract segmentation using a tractography-based atlas method \citep{de2015tract}. The method defined standard space atlases that were non-linearly transformed to subject native space. These atlases guided probabilistic tractography, which was performed with its default settings in FSL (PROBTRACKX; diffusion model was estimated using BEDPOSTX) \citep{jenkinson2012fsl}. Tractography protocols are available as the FSL AutoPTX\footnote{\url{http://fsl.fmrib.ox.ac.uk/fsl/fslwiki/AutoPtx}} plugin \citep{de2013improving}. The resulting tract-specific density images were normalized by division with the total number of tracts in the tractography run. Finally, tract-specific thresholds were established by maximizing the FA reproducibility on a training set of 30 subjects with 2 scans. Volume-based tract outliers were visually inspected. We excluded all scans for which one or more tracts did not pass quality control \citep{de2015tract}. 

\subsubsection{White matter tract segmentation model}
\label{sssec:2.1.6}
We propose a direct WM tract segmentation model that takes a 4D diffusion tensor image as input. Let $I_{dti}\in R^{i\times j\times k\times6} $ denote a tensor image in native diffusion space, and $I_{seg}\in R^{i\times j\times k} $ denote the reference segmentation of a WM tract. The segmentation process finds a relation

\begin{equation}
 I_{seg} = \mathcal{F}\boldsymbol{_\Theta} (I_{dti}),
\end{equation} 
which is parameterized by $\boldsymbol{\Theta}$. Then $\boldsymbol{\Theta}$ can be optimized by minimizing the loss function $\mathcal{L}$:

\begin{equation}
 \argmin_{\boldsymbol{\Theta}}  \mathcal{L}\big(\mathcal{F}\boldsymbol{_\Theta} (I_{dti}),I_{seg}\big).
\end{equation}

\begin{figure*}[!ht]
  \centering
   \includegraphics[scale=0.46]{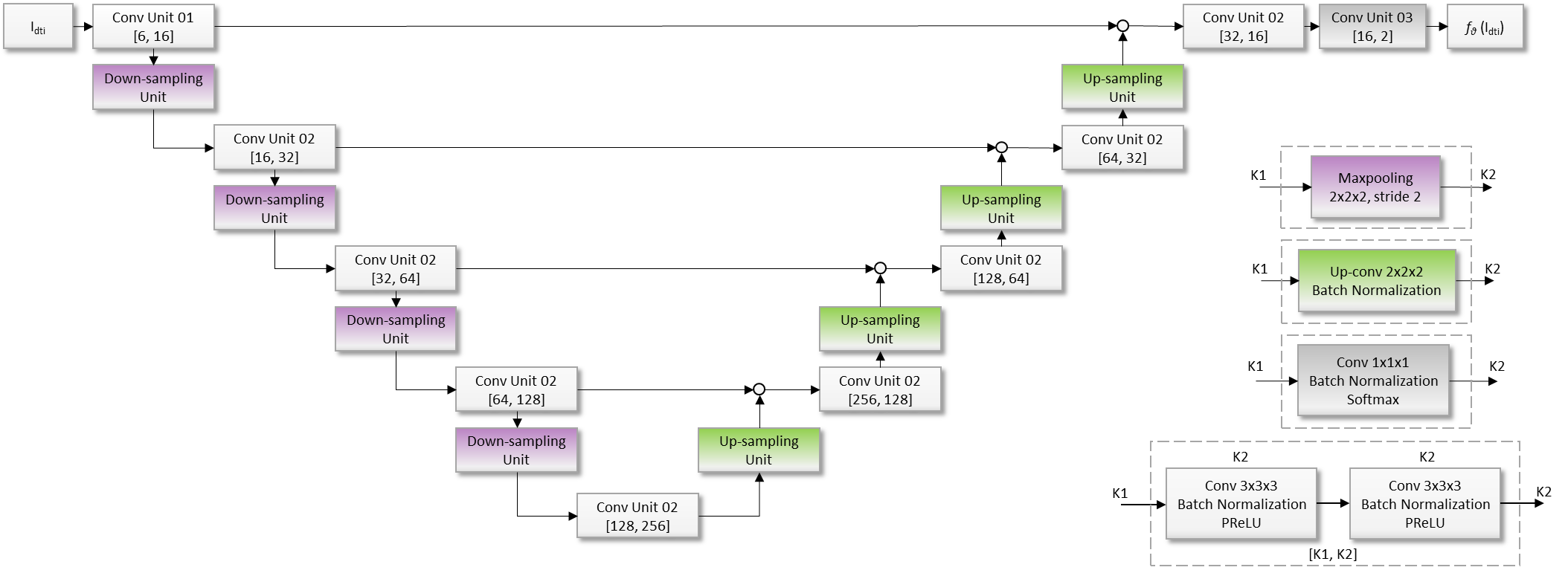}
  \caption{The proposed 3D-CNN encoder-decoder architecture for WM tract segmentation. The colored boxes in the lower right corner detail corresponding units in the network architecture, where $[k_1, k_2]$ are the number of convolution kernels in those layers. Abbreviations: Conv = convolution, PRelu = parametric rectified linear units. The circles in the decoder path indicate concatenation operations.}\label{fig1}
\end{figure*}

The relation $\mathcal{F}\boldsymbol{_\Theta}$ is modeled by a 3D CNN, which consists of a series of convolutions and non-linearity operations. An encoder-decoder network (Figure \ref{fig1}) is used according to the U-Net architecture \citep{ronneberger2015u} with additional skip connections. The encoder path is a gradual compression process of extracting abstract features from the diffusion tensor images, in which all but the maximum values within a kernel were discarded after each max-pooling layer. Then the decoder path restores the details and combines them with the shallow information of the same scales. The convolution layers produce a set of $k$ feature maps by individually convolving the input with $k$ kernels. The size of the convolution kernels in the last layer were  $1\times1\times1$, those in the other layers were $3\times3\times3$. For parameter regularization and accelerating model training, convolution layers were followed by batch normalization \citep{ioffe2015batch}. Non-linearities were defined using parametric rectified linear units (PReLU) \citep{he2015delving}. The last layer of the network was a voxelwise softmax function that outputs a probability map $P(I_{seg}|\boldsymbol{\Theta},I_{dti})$. For performance evaluation, probabilistic segmentations were binarized ($P>0.5$).

A separate model was trained for each tract. In each training epoch, input volumes were fed in random batches (size $=2$) for robustness. To improve efficiency, batches were generated ``on-the-fly''. We used the Adam optimizer \citep{kingma2014adam} with an initial learning rate of 0.1, which was adaptively reduced by 50\% once the validation loss stopped improving for 15 epochs. For tracts that are left/right homologous, the combined dataset was used for pre-training. 

\subsubsection{Evaluation metric}
\label{sssec:2.1.7}
Segmentation accuracy was quantified by the Dice coefficient between the segmentation result $\big(\mathcal{F}\boldsymbol{_\Theta} (I_{dti})\big)$ and the reference segmentation ($I_{seg}$). The dice coefficient (DC) was computed within the bounding box ROI and followed its definition:
\begin{equation}
DC\big(\mathcal{F}\boldsymbol{_\Theta} (I_{dti}),I_{seg}\big)=\frac{2 \times |\mathcal{F}\boldsymbol{_\Theta} (I_{dti})\cap I_{seg}|}{|\mathcal{F}\boldsymbol{_\Theta} (I_{dti})|+|I_{seg}|},
\end{equation}
where $|.|$ is the cardinality. 

\subsection{Optimization experiments}
\label{ssec:2.2}

We optimized the method on three key elements: 1) input, 2) network architecture, and 3) the loss function and tract weight. The following sections describe the optimization experiments, for which the forceps minor (FMI) tract was used. This tract was chosen since it has previously shown importance in neurodegeneration and aging \citep{rascovsky2011sensitivity} and is relatively complex to segment due to thin structure. Paired sample t-tests ($\alpha=0.05$) and Bonferroni correction for controlling the family-wise error of multiple testing were used to test the statistical significance of comparisons. Experiments were performed on one node of the Dutch national supercomputer Cartesius which consists of Intel E5-2450 v2 CPUs and NVidia Tesla K40m GPUs.

\subsubsection{Experiment 1: input}
\label{sssec:2.2.1}
As method inputs, we evaluated the T1-weighted image (\textit{T1w}) as well as several dMRI-based images, i.e., the diffusion tensor image (\textit{tensor}), and the FA and MD image (\textit{FA + MD}). Because of prior knowledge, \textit{tensor} was always included as input: tensor implicitly contains information on crossing fibers and can be decomposed into other diffusion measure images. To assess the added value of spatial information, we additionally evaluated an input image encoding \textit{location}. The \textit{location} data includes voxel-wise coordinates that map each diffusion volume to the T1 MNI152 image \citep{evans19933d}. These coordinates were obtained by non-linear transformation of the coordinates of the MNI152 image to the subject native T1w space, concatenated with a linear transformation to the subject native diffusion space using FNIRT and FLIRT \citep{jenkinson2002improved}. Using the proposed network architecture and weighted inner product loss function ($W=3$), we trained models on eight different combinations of inputs: 1. \textit{tensor}, 2. \textit{tensor + T1w}, 3. \textit{tensor + FA + MD}, 4. \textit{tensor + FA + MD + T1w}, 5. \textit{tensor + location}, 6. \textit{tensor + location + T1w}, 7. \textit{tensor + FA + MD + location}, and 8. \textit{tensor + FA + MD + location + T1w}. Correcting for 7 tests resulted in an adjusted P-value threshold of $7.1 \times 10^{-3}$.

\begin{figure*}[!ht]
  \centering
   \subfigure[Input selection]{
   \includegraphics[scale=0.45]{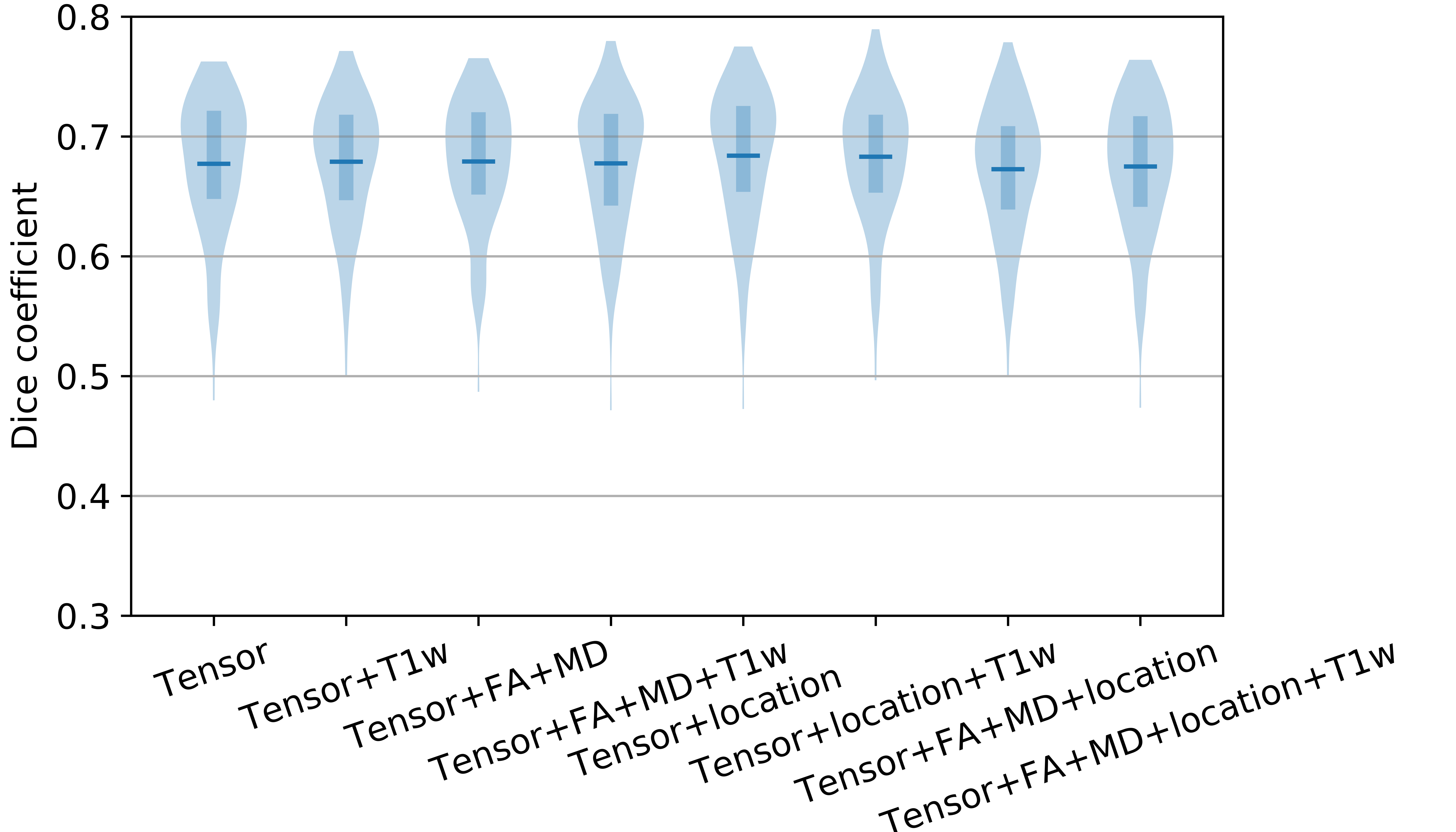}}
   \qquad
   \subfigure[Architectures and optimizers]{
   \includegraphics[scale=0.45]{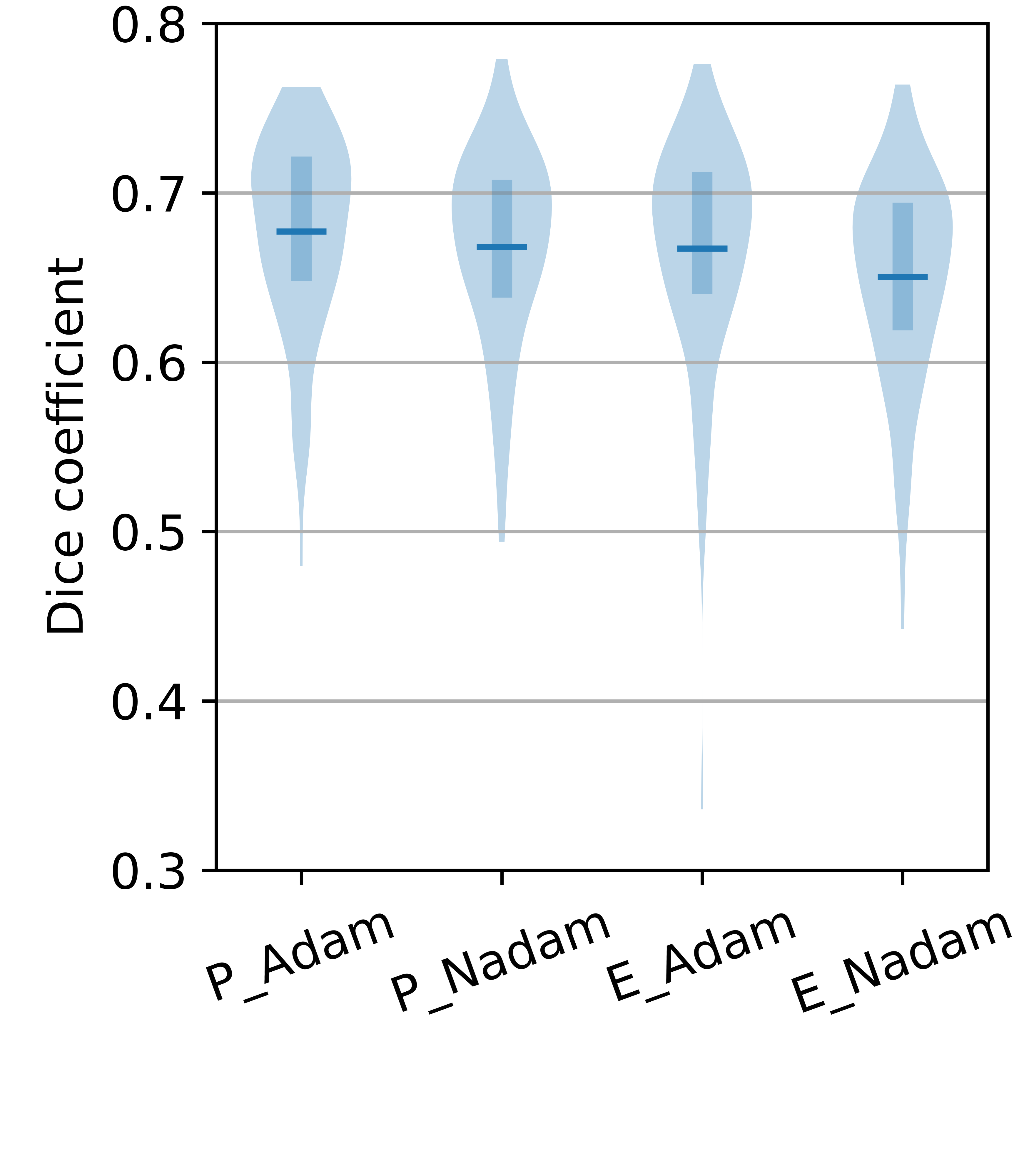}}
  \caption{FMI segmentation accuracy (DC) on $D1_{test}$ using different (a) model inputs (Exp. 1), and (b) architectures and optimizers (Exp. 2). \textit{Location}: input image describing spatial information, \textit{P}: proposed architecture, \textit{E}: {\itshape Ext-architecture}. In the violin plots, horizontal lines refer to the mean, and vertical lines refer to the range of the first quartile and the third quartile.}\label{fig2}
\end{figure*}

Results are presented in Figure~\ref{fig2} (a). All combinations showed a similar accuracy with a mean DC of $0.68$. The additional features (\textit{FA + MD, location, T1w}) did not improve significantly the model based on \textit{tensor} only ($p>0.007$). Hence, the model based on \textit{tensor} was optimal in this setting. 

Additionally, we evaluated the method using the first three peaks of the fiber orientation distribution function (fODF) as input. The fODF peaks were estimated with the single-shell single-tissue setting of the Constrained Spherical Deconvolution function, available in MRtrix \citep{tournier2007robust}. The test DC of the model trained on fODF peaks using the proposed network architecture and weighted inner product loss function ($W=3$) is $0.41 \pm 0.17$. This fODF-based performance was significantly lower than the tensor-based performance ($p=5.8\times 10^{-41}$).

\subsubsection{Experiment 2: network architecture}
\label{sssec:2.2.2}
We compared the proposed architecture (Section~\ref{sssec:2.1.6}, Figure~\ref{fig1}) with an extended architecture, {\itshape Ext-architecture} (Supplementary Figure~\ref{Sfig1}). The {\itshape Ext-architecture} is an extension of the proposed architecture with the addition of novel convolutional re-samplings and a residual function. In short, the max-pooling operation was replaced by strided convolution, and the up-sampling was replaced by convolution transpose. This introduces trainable parameters which allow the network to explore the way of re-sampling itself. Also, the residual function used in {\itshape Ext-architecture} adds the input to the output of each convolution layer, which is processed through the convolution and non-linearities, to reformulate feature representation between a finer and a coarser scale. This was expected to improve segmentation accuracy \citep{milletari2016v}.

In addition, we compared two gradient descent algorithms (with default parameters) for our setting: Adam and Nadam \citep{dozat2016incorporating}. The models were trained on the \textit{tensor} input using weighted inner product loss function ($W=3$). Correcting for 3 tests resulted in an adjusted P-value threshold of $1.7 \times 10^{-2}$.

Figure~\ref{fig2} (b) shows the test DC of the two architectures in combination with the different optimizers. For both optimizers, the proposed network architecture yielded statistically significantly ($p<0.01$) a higher segmentation accuracy than the {\itshape Ext-architecture} ($P_{Adam}=0.68$, $E_{Adam}=0.67$) and a lower standard deviation ($P_{Adam}=0.054$, $E_{Adam}=0.065$). The Adam optimizer ($p<0.01$).

\subsubsection{Experiment 3: loss function and tract weight}
\label{sssec:2.2.3}
We propose to use the weighted inner product ($\mathcal{L}_{wip}$) \citep{choi2010survey} as a loss function:  

\begin{equation}\label{eq:4}
\mathcal{L}_{wip} = - W \times I_{seg} \times \mathcal{F}\boldsymbol{_\Theta} (I_{dti}) - \Big(1 - I_{seg}\Big) \times \Big(1 - \mathcal{F}\boldsymbol{_\Theta} (I_{dti})\Big), 
\end{equation}
where $W$ is the weight of the tract class.

We compared its performance to that of the widely used weighted cross entropy ($\mathcal{L}_{wce}$) loss function in our setting (i.e., \textit{tensor} input, proposed network architecture and the Adam optimizer), which is defined as: 
\begin{equation}
\mathcal{L}_{wce} = - W \times I_{seg} \times log\Big( \mathcal{F}\boldsymbol{_\Theta} (I_{dti})\Big) - \Big(1 - I_{seg}\Big) \times log\Big(1 - \mathcal{F}\boldsymbol{_\Theta} (I_{dti})\Big). 
\end{equation}

The tract weight trades off between recall and precision of the segmentation. To tune the tract weight and balance classes, we evaluated different $W$ ranging from 0.5 to the mean frequency ratio of non-tract and the tract voxels ($W=100$). Correcting for 11 tests resulted in an adjusted P-value threshold of $4.5 \times 10^{-3}$.

\begin{figure}[!ht]
  \centering
   \includegraphics[scale=0.45]{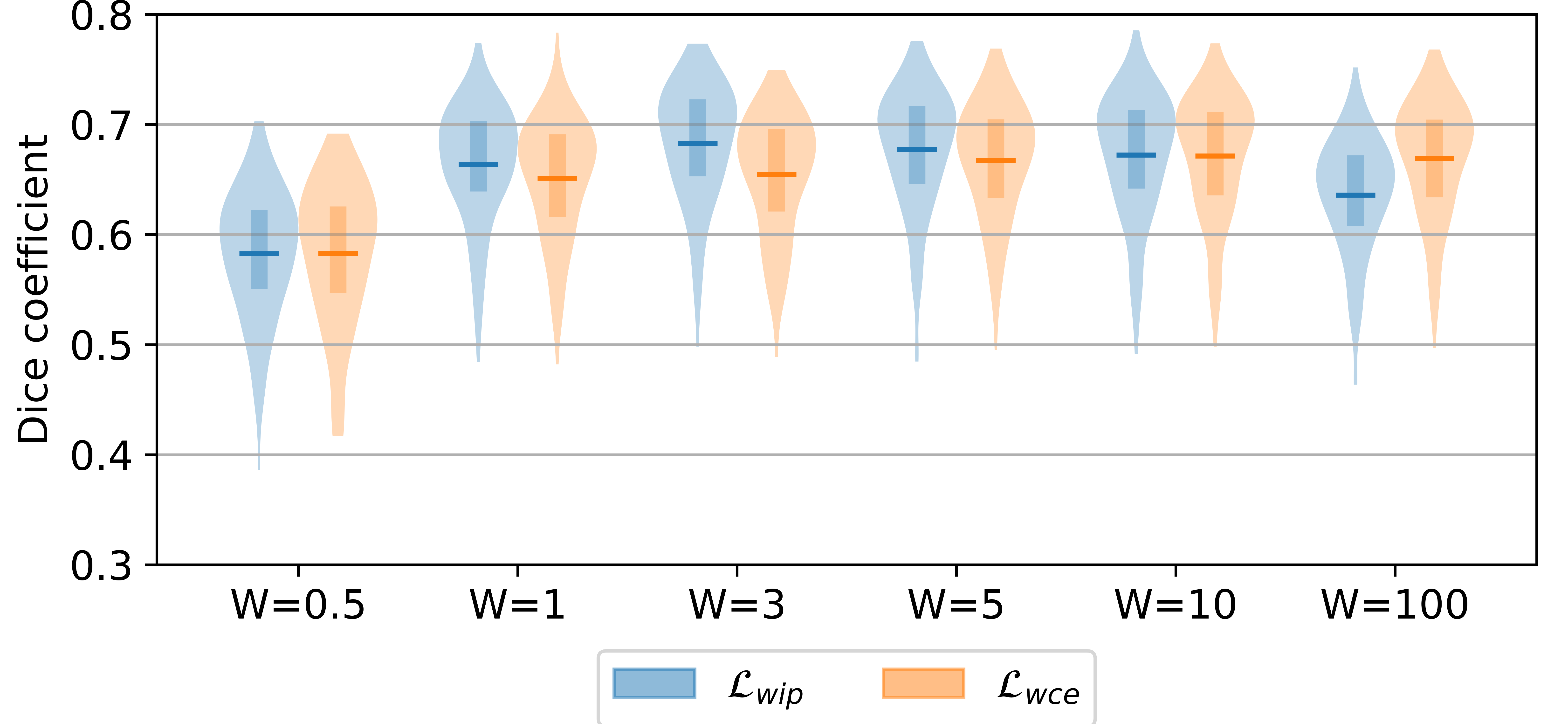}
  \caption{FMI segmentation accuracy (DC) on $D1_{test}$ using $\mathcal{L}_{wip}$ and $\mathcal{L}_{wce}$ loss functions. $W$ indicates the tract weight. In the violin plots, horizontal lines refer to the mean, and vertical lines refer to the range of the first quartile and the third quartile.}\label{fig4}
\end{figure}

The results obtained using $\mathcal{L}_{wip}$ and $\mathcal{L}_{wce}$ loss functions and 6 tract weights are provided in Figure \ref{fig4}. For both loss functions, a weight between 1 and 10 gave relatively optimal performance. The highest DC was achieved using $\mathcal{L}_{wip}$ at $W=3$, although the differences in DC with $W=1$ and $W=5$ were not statistically significant ($p>0.005$). Overall, the $\mathcal{L}_{wip}$ performed better than the $\mathcal{L}_{wce}$ in this setting. Comparing with using default cross-entropy loss function ($\mathcal{L}_{wce}, W=1$), the use of proposed loss function in combination with optimal tract weight ($\mathcal{L}_{wip}, W=3$) significantly improved the accuracy (DC) from $0.65\pm0.06$ to $0.68\pm0.05$ ($p<0.001$). This is also significantly better than performance obtained with the optimal weight ($W=5$) for the $\mathcal{L}_{wce}$ loss ($p<0.001$).

\subsubsection{Optimization results}
Neuro4Neuro adopted the best settings of the three optimization experiments: the diffusion tensor elements as input, the proposed network architecture (Figure~\ref{fig1}), and the weighted inner product loss function ($L_{wip},W=3$) with the Adam optimizer. 

We compared the performance of our method to a basic atlas-based segmentation pipeline. Specifically, a probabilistic tract heatmap was established by non-linearly co-registering the reference segmentations of all training data to the FMRIB58\_FA\_1mm template using FLIRT, FNIRT, and the FA\_2\_FMRIB58\_1mm protocol \citep{jenkinson2002improved}. The normalized probabilistic atlas was then registered to each test image using the same protocol and binarized with a threshold of 0.5. The averaged DC over test dataset is $0.44\pm0.08$, significantly lower than that of the optimized setting for Neuro4Neuro ($p<0.01$).

\section{Validation on a normal and a dementia population}
\label{s:3}

\subsection{Materials}
\label{ssec:3.1}
\subsubsection{Study population}

\label{sssec:3.1.1}

The normal population consisted of community-dwelling elderly from the Rotterdam Study (Section~\ref{sssec:2.1.1}). Their imaging data were split into several subsets: a training set ($D2_{train}$) consisting of 7079 scans from 3858 participants (including the optimization set $D1$), a test set ($D2_{test}$) consisting of 1104 scans from 1104 participants, and an additional set for testing reproducibility ($D3$) consisting of 194 scans from 97 participants. The participants in $D3$ had been scanned twice with a mean interval of 20.2 days. We ensured that the testing sets, i.e., $D2_{test}$ and $D3$, did not contain any scans of participants in the training set.  

The dementia population ($D4$) consisted of behavioural variant frontotemporal dementia (bvFTD) patients, Alzheimer's disease (AD) patients, and cognitively healthy participants from the Iris study \citep{steketee2016early}. The Iris study was approved by the local medical ethics committee. All participants gave written informed consent. MRI scans of the patients were obtained at baseline and at one year ($383.9 \pm 9.9$ days) follow-up; controls were scanned at baseline only. After quality control, twelve bvFTD patients, eleven AD patients and eighteen controls were included in our analysis (Table \ref{tab1}) \citep{meijboom2019exploring}. 

\begin{table}
\centering
\caption{Demographic characteristics of $D4$, adapted from \cite{meijboom2019exploring}. N is the sample size.  SD: standard deviation, bvFTD: behavioural variant frontotemporal dementia, AD: Alzheimer's disease, T0: baseline, T1: one-year follow-up, MMSE: mini-mental state examination score.}\label{tab1}
\begin{tabular*}{0.48\textwidth}{cccc}
 \hline
Group&N (male)&Mean age (SD)&Mean MMSE\\
 \hline
BvFTD, $T0$ &12 (6) &60.3 (7.7)&26.6 (2.8)\\
BvFTD, $T1$ &6 (3) &64.0 (3.6)&-\\
AD, $T0$ &11 (8) &62.8 (5.0)&25.3 (2.0)\\
AD, $T1$ &11 (8) &63.3 (5.0)&-\\
Controls, $T0$ &18 (8) &59.8 (6.7)&29.1 (1.0)\\
\hline
\end{tabular*}
\end{table}

\subsubsection{MRI acquisition}
\label{sssec:3.1.2}
For the Rotterdam Study, the MRI protocol is described in Section \ref{sssec:2.1.2}. 
For the Iris study, scans were acquired on a 3T MRI scanner (GE Discovery MR750). The acquisition parameters of the diffusion images were: $TR/TE = 7930 ms/84.5 ms$; imaging matrix of $128\times128$ in an FOV of $240\times240 mm^2$; 25 diffusion weighted volumes ($b=1000 s/mm^2$) and three non-weighted volumes ($b=0 s/mm^2$). The voxel size was $1.8\times1.8\times2.5 mm^3$.

\subsubsection{Image preprocessing}
\label{sssec:3.1.4}

DWI data were corrected for motion and eddy currents using the pipeline described in Section \ref{sssec:2.1.3}. For the Iris study, scans were subsequently resized to match the same image size with the Rotterdam Study data, diffusion tensors and measures were estimated using DTIFIT \citep{behrens2003characterization}. Tract-specific measures were computed as the mean value of non-zero diffusion measures within each segmented tract.

\subsection{Experiments}
\label{ssec:3.2}
Experiments were performed to assess general performance in terms of accuracy, reproducibility and generalizability. In addition, we performed proof-of-concept clinical application experiments: 1) the association between age and diffusion measures in normal aging and 2) differential diagnosis of bvFTD and AD. For these experiments, we trained the optimized model from Section \ref{ssec:2.1} on $D2_{train}$ for 25 tracts of four categories: 1) the Association tracts: anterior thalamic radiation (ATR), inferior fronto-occipital fasciculus (IFO), inferior longitudinal fasciculus (ILF), posterior thalamic radiation (PTR), superior longitudinal fasciculus (SLF) and uncinate fasciculus (UNC); 2) the Commmissural tracts: forceps major (FMA) and forceps minor (FMI); 3) the Limbic tracts: cingulate gyrus part of cingulum (CGC) and parahippocampal part of cingulum (CGH); and 4) the Sensorimotor tracts: corticospinal tract (CST), middle cerebellar peduncle (MCP), medial lemniscus (ML) and superior thalamic radiation (STR). Analyses were performed using Python 3.6.3 (SciPy and Sklearn package) and SPSS (version 24). 

\subsubsection{Accuracy}
\label{sssec:3.2.1}
Segmentation accuracy was measured in $D2_{test}$ using the tract-specific DC between the model's binary segmentation and the reference segmentation. 

\subsubsection{Reproducibility}
\label{sssec:3.2.2}
The reproducibility of the proposed method was evaluated statistically both based on diffusion measures and tract volume, and based on voxel-wise agreement of the segmentations. For these experiments, each scan in $D3$ was segmented separately. Because of the short time interval between the two scans of each participant, the tract segmentations, volumes, and diffusion metrics are expected to be identical. 

For the reproducibility of tract-specific diffusion measures and volumes, we quantified these values in their native space and computed the relative difference in paired scan-rescan measures ($m_1, m_2$) as an indicator of error ($\epsilon$), which was defined as
\begin{equation}
\epsilon =\frac{\lvert m_{2} - m_{1} \lvert }{\frac{1}{2} (m_{2} + m_{1})} \times 100 \%\label{equ6}.
\end{equation}
A lower $\epsilon$ indicates a better reproducibility. The $R^2$ values of ordinary least squares regression for tract-specific FA, MD and volume were also computed. A higher $R^2$ value indicates a better reproducibility.
\begin{figure*}[!ht]
  \centering
     \subfigure[Dice coefficient]{
  	 \includegraphics[scale=0.46]{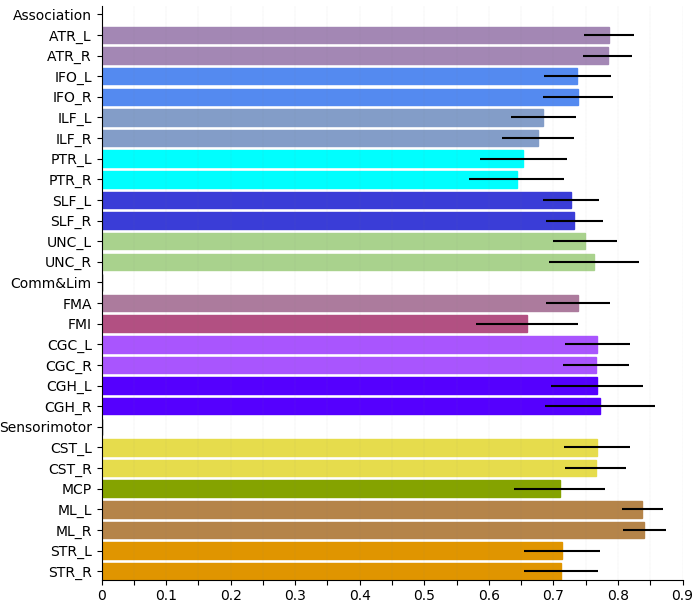}}
  	 \quad
     \subfigure[Result visualization]{
  	 \includegraphics[scale=0.29]{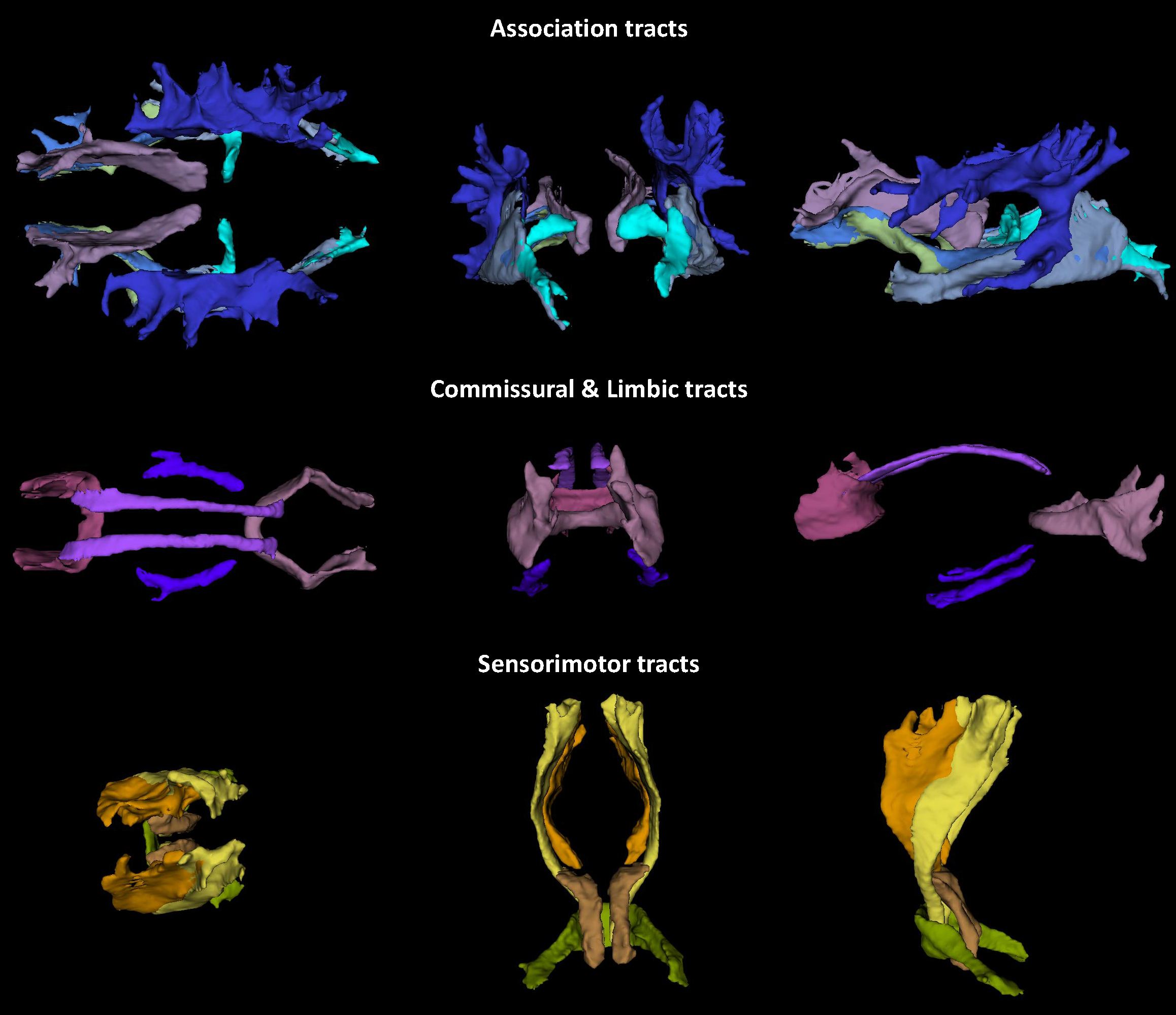}}
 	 \caption{Test results of 25 tracts on $D2_{test}$. (a) Dice coefficients between Neuro4Neuro predictions and the reference segmentation. ``Comm\&Lim'': the Commissural and Limbic tracts. (b) Individual tracts of a participant (75 years old, female) analyzed by Neuro4Neuro, showed in association (top row), commissural and limbic (middle row), and sensorimotor (bottom row) tract groups. In each row, superior, posterior and left views are shown. ATR: anterior thalamic radiation, CGC: cingulate gyrus part of cingulum, CGH: parahippocampal part of cingulum, CST: corticospinal tract, FMA: forceps major, FMI: forceps minor, MCP: middle cerebellar peduncle, ML: medial lemniscus, IFO: inferior fronto-occipital fasciculus, ILF: inferior longitudinal fasciculus, PTR: posterior thalamic radiation,  SLF: superior longitudinal fasciculus,  STR: superior thalamic radiation, UNC: uncinate fasciculus.}\label{fig5}
\end{figure*}

For quantifying reproducibility in terms of voxel-wise agreement between the segmentations, we used the Cohen's kappa ($\kappa$) coefficient. Typically, a $\kappa > 0.60$ indicates ``substantial'' agreement, and a $\kappa > 0.80$ indicates ``almost perfect'' agreement \citep{landis1977measurement}. The segmentations ($s_{1}, s_{2}$) of two scans were obtained independently and subsequently aligned based on rigid registration of the corresponding FA images using Elastix \citep{klein2010elastix}. $\kappa$ is defined as 
\begin{equation}
\kappa =\frac{p_o - p_e}{1-p_e}\label{equ7},
\end{equation}
in which $p_o$ is the observed agreement between $s_{1}$ and $s_{2}$, and $p_e$ is the hypothetical probability of the agreement. Given N is the total number of voxels in the scan, $n_{t,s}$ is the number of voxels in a segmentation that is predicted as a specific tract, and $n_{n,s}$ is the number of background voxels (non-tract), the hypothetical probability of the agreement can be estimated by
\begin{equation}
p_e=\frac{1}{N^2}(n_{t,s_{1}}\times n_{t,s_{2}}+n_{n,s_{1}}\times n_{n,s_{2}}).
\end{equation}

Paired sample t-tests ($\alpha=0.05$) were used to test the statistical significance of above metrics in comparison with those of the reference method. 

\subsubsection{Application in normal aging}
\label{sssec:3.2.3}

We evaluated the applicability of our method to study tract-specific measures by replicating a population-based analysis of neurodegeneration in aging. This statistical analysis was performed in the $D2_{test}$ sample according to a approach adapted from \cite{de2015tract}.

In short, we associated microstructural diffusion measures with aging using multi-variable linear regressions. For left/right homologous tracts, we computed the volume-wise average of the tract-specific measures, i.e., FA, MD, L1, RD and MO.  Two regression models with different confounding regressors were fitted for each tract. Analyses were adjusted for sex and intracranial volume (ICV)  (Model 1). Supratentorial ICV was estimated by summing total WM, grey matter and cerebrospinal fluid volumes. Additionally, we adjusted for tract-specific volume in the Model 2. An $\alpha=0.05$ and Bonferroni correction for controlling the family-wise error of multiple testings were used. Taking into account the three eigenvalues tested in five compositions (FA, MD, and three additional diffusion measures), correcting for 84 tests (28 models, 3 eigenvalues) resulted in an adjusted P-value threshold of $6.0\times10^{-4}$. Analyses were performed using both the proposed method and the reference method.

\subsubsection{Generalizability and application in dementia}
\label{sssec:3.2.4}

The proposed method was assessed for generalizability to an external dataset and for its value in groupwise differentiation of dementia at multiple time-points. For this experiment, the method was trained on the data of normal population $D2_{train}$ and tested on the dementia data $D4$. 
Generalizability was assessed qualitatively by comparing the segmentation with those obtained by the reference method  (Section \ref{sssec:2.1.5}). Because of resolution differences, we adjusted the tract-specific threshold of the reference method \citep{meijboom2019exploring}.

In addition, we evaluated the applicability of Neuro4Neuro for studying neurodegenerative diseases by replicating an analysis of differentiating early-stage dementia based on tract-specific measures. This statistical analysis was performed in the $D4$ sample using an approach adopted from \cite{meijboom2019exploring}, in which the reference segmentation method was utilized for tracts segmentation. Sixteen tracts were included in the analysis, excluding the left and right PTR, CST, ML and STR, and MCP tracts. Tract-specific diffusion measures (i.e., FA, MD, L1 and RD) at baseline were group-wise compared among bvFTD, AD and controls using ANOVA and post-hoc Bonferroni t-test. In case of unequal variances across groups, a Welch-ANOVA and post-hoc Games-Howell t-tests were used. Also, the same approach was used to analyze diffusion measures at follow-up between the bvFTD and AD groups. 

\subsection{Results}
\label{ssec:3.3}

\begin{figure*}[!ht]
     \subfigure[]{
  	 \includegraphics[scale=0.49]{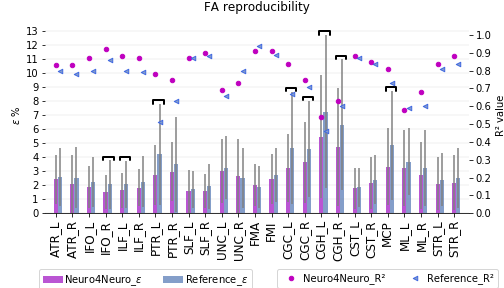}}
  	 \quad
     \subfigure[]{
  	 \includegraphics[scale=0.49]{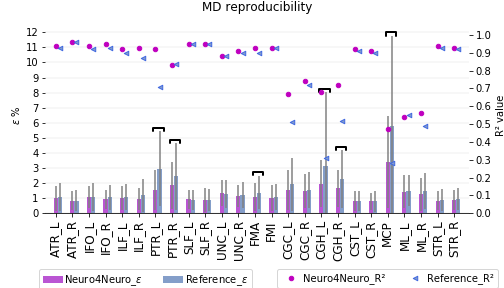}}     

	 \subfigure[]{
	 \includegraphics[scale=0.49]{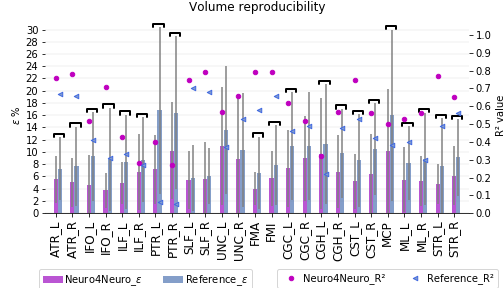}}
  	 \quad
     \subfigure[]{
  	 \includegraphics[scale=0.49]{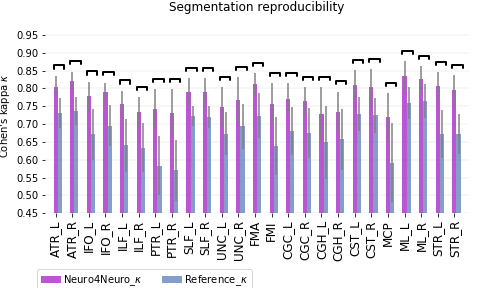}}
 	 \caption{Reproducibility of tract-specific measures over $D_3$. FA: fractional anisotropy, MD: mean diffusivity ($10^{-3} mm^{2}/s$), Volume: tract volume (ml). $\epsilon$ indicates relative scan-rescan difference in measures (Eq.~\ref{equ6}). A lower $\epsilon$ indicates a better reproducibility. Error bars indicate standard deviations. $R^2$ value was obtained by OLS regression for scan-rescan measures. A higher $R^2$ value indicates a better reproducibility. $\kappa$, the Cohen's kappa coefficient, indicates spatial correspondence of the segmentation (Eq.~\ref{equ7}). A higher $\kappa$ indicates a better reproducibility. The bold bracket indicates that the $\epsilon$ was significantly lower or the $\kappa$ was significantly higher for Neuro4Neuro than for the reference method (t-test, $p<0.05$). }\label{fig6}
\end{figure*}

\subsubsection{Accuracy}
\label{sssec:3.3.1}

Figure~\ref{fig5} shows the segmentation accuracy on $D2_{test}$ and an example visualization of the tract segmentations. For visualization, we selected a participant whose DC was equal to the mean value on $D2_{test}$. The mean accuracy over 25 WM tracts was $DC= 0.74$ (range: $0.64-0.84$). 

\subsubsection{Reproducibility}
\label{sssec:3.3.2}
The reproducibility of tract-specific FA, MD, volume, and segmentation is shown in Figure~\ref{fig6}. The proposed method overall led to higher reproducibility than the reference method, i.e., lower errors in scan-rescan measures ($\epsilon$), higher $R^2$ values, and higher spatial correspondence ($\kappa$). The difference in the average $\epsilon$ between the two methods was statistically significant ($p<0.05$) in 6 tracts for MD, in 8 tracts for FA, and in 20 tracts for volume. Among 25 tracts, the $\epsilon$ of our method was lowest for the MD measures (mean = $1\%$, range: $1\% - 3\%$), followed by FA measures (mean = $3\%$, range: $1\% - 5\%$) and volume measures (mean = $7\%$, range: $4\% - 11\%$). Those for the reference method were: MD (mean = $2\%$, range: $1\% - 6\%$), FA (mean = $3\%$, range: $2\% - 7\%$), and volume (mean = $10\%$, range: $6\% - 17\%$).

The $R^2$ values of tract-specific measures were generally higher for our method, especially for the volume metric when comparing with those of the reference method. For the proposed method, the $R^2$ value over 25 tracts was highest for the MD measures (mean=$0.84$, range: $0.47 - 0.96$), followed by FA  (mean=$0.80$, range: $0.54 - 0.92$) and volume (mean=$0.59$, range: $0.27 - 0.79$). Those for the reference method were: MD (mean=$0.78$, range: $0.28 - 0.96$), FA (mean=$0.75$, range: $0.46 - 0.94$), and volume (mean=$0.44$, range: $0.05 - 0.70$).

In addition, the segmentations of scan-rescan data analyzed by our method showed a ``substantial'' to ``almost perfect'' spatial correspondence ($\kappa$, mean=$0.78$, range: $0.72 - 0.83$) as seen in Figure~\ref{fig6}(d). The difference in Cohen's kappa ($\kappa$) between two methods was significant for all 25 tracts. $\kappa$ for the reference method was: (mean=$0.68$, range: $0.57 - 0.76$).

\subsubsection{Tract-specific neurodegeneration in aging}
\label{sssec:3.3.3}

\begin{table*}[!ht]
 \scriptsize
\caption{Associations between age and tract-specific diffusion measures. Values ($\times 10^{-3}$) represent regression coefficients ($\beta$) and their standard error ($std.error$) for change in fractional anisotropy (FA) or mean diffusivity (MD) per year increase in age, adjusted for sex and ICV (and additionally for tract-specific WM volume in Model 2). Significant associations at Bonferroni corrected threshold $P-value=6.0\times10^{-4}$ are shown in \textbf{bold}.}.\label{tab2}

\begin{tabular*}{0.99\textwidth}
{@{}@{\extracolsep{\fill}}cccccccccc@{}}
\hline
\multicolumn{1}{c}{ }  &
\multicolumn{4}{c}{Model 1} &
\multicolumn{1}{c}{}   &
\multicolumn{4}{c}{\makecell{Model 2: \\Model 1 + WM volume}}  \\

\cline{2-5}\cline{7-10}
 Tract &FA $\beta$ & FA $std.error$  &MD $\beta$& MD $std.error$ & &FA $\beta$ &FA $std.error$ &MD $\beta$&MD $std.error$\\
 \hline
{\itshape Association} \\ 
{ATR}    \\
{Neuro4Neuro} &\textbf{-1.05} &0.10 &\textbf{4.90}  &0.25 & & \textbf{-0.71} &0.10&\textbf{3.83}&0.25  \\
{Reference} &\textbf{-1.09} &0.10 &\textbf{4.88}  &0.24 & & \textbf{-0.85} &0.10&\textbf{4.17}&0.24  \\
{IFO}    \\
{Neuro4Neuro} &\textbf{-1.75} &0.12 &\textbf{4.38}  &0.22 & & \textbf{-1.31} &0.13&\textbf{3.71}&0.23  \\
{Reference}   &\textbf{-1.81} &0.13 &\textbf{4.10}  &0.22 & & \textbf{-1.65} &0.12&\textbf{3.99}&0.22 \\

{ILF}    \\
{Neuro4Neuro} &\textbf{-0.93} &0.12 &\textbf{3.09}  &0.20 & & \textbf{-0.82} &0.12&\textbf{3.02}&0.20 \\
{Reference}   &\textbf{-1.01} &0.12 &\textbf{3.18}  &0.21 & & \textbf{-1.00} &0.12&\textbf{3.17}&0.21  \\

{PTR}    \\
{Neuro4Neuro}  &\textbf{-1.48} &0.11 &\textbf{5.34}  &0.29 & & \textbf{-1.46} &0.12&\textbf{5.39}&0.29 \\
{Reference}    &\textbf{-1.41} &0.13 &\textbf{5.02}  &0.32 & & \textbf{-1.36} &0.12&\textbf{5.09}&0.32 \\

{SLF}    \\
{Neuro4Neuro}  &\textbf{-0.74} &0.12 &\textbf{2.15}  &0.19 & & \textbf{-0.70} &0.12&\textbf{2.11}&0.19 \\
{Reference}   &\textbf{-0.89} &0.12 &\textbf{2.13}  &0.19 & & \textbf{-0.94} &0.11&\textbf{2.17}&0.19  \\

{UNC}    \\
{Neuro4Neuro} &\textbf{-1.40} &0.11 &\textbf{2.83}  &0.16 & & \textbf{-1.15} &0.11&\textbf{2.70}&0.16  \\
{Reference}  &\textbf{-1.43} &0.12 &\textbf{2.85}  &0.16 & & \textbf{-1.22} &0.10&\textbf{2.78}&0.16   \\

{\itshape Commissural} \\
{FMA}    \\
{Neuro4Neuro} &\textbf{-2.01} &0.17& \textbf{3.27}& 0.27& & \textbf{-1.01}&0.15 &\textbf{2.34} &0.27  \\
{Reference}   &\textbf{-2.36} &0.18& \textbf{3.36}& 0.29& & \textbf{-1.67}&0.17 &\textbf{2.84} &0.29\\

{FMI}     \\
{Neuro4Neuro}   &\textbf{-2.64} &0.17& \textbf{2.93}& 0.19& & \textbf{-1.28}&0.16 &\textbf{2.14} &0.20  \\
{Reference}   &\textbf{-2.69} &0.18& \textbf{2.83}& 0.20& & \textbf{-1.91}&0.16 &\textbf{2.54} &0.20  \\

{\itshape Limbic} \\
{CGC}    \\
{Neuro4Neuro} &\textbf{-1.48}&0.19&\textbf{1.09}&0.13&  &\textbf{-1.34}&0.17&\textbf{1.06}&0.13  \\
{Reference}  &\textbf{-1.51}&0.19&\textbf{1.08}&0.13&  &\textbf{-1.59}&0.19&\textbf{1.09}&0.13  \\

{CGH}     \\
{Neuro4Neuro} &\textbf{-1.18} &0.13& \textbf{1.86}& 0.18& & \textbf{-1.10}&0.13 &\textbf{1.82} &0.18 \\
{Reference}    &\textbf{-1.32} &0.14& \textbf{1.95}& 0.22& & \textbf{-1.31}&0.14 &\textbf{1.94} &0.22 \\

{\itshape Sensorimotor} \\
{CST}    \\
{Neuro4Neuro} &-0.07 &0.13& \textbf{1.59}& 0.14& & -0.03&0.13 &\textbf{1.39} &0.15  \\
{Reference}  &-0.29 &0.13& \textbf{1.77}& 0.14& & -0.12&0.13 &\textbf{1.51} &0.14  \\

{MCP}   \\
{Neuro4Neuro} &-0.27 &0.20& 0.63& 0.40& & -0.35&0.20 &0.04 &0.35 \\
{Reference}   &-0.49 &0.22& 0.51& 0.48& & -0.62&0.22 &-0.15 &0.42 \\

{ML}     \\
{Neuro4Neuro}  &0.22 &0.11& 0.17& 0.11& &0.30&0.11 &0.17 &0.11 \\
{Reference}    &0.05 &0.11& 0.26& 0.13& &0.11&0.11 &0.27 &0.13 \\

{STR}     \\
{Neuro4Neuro} &-0.44 &0.13& \textbf{2.46}& 0.17& &-0.19&0.13 &\textbf{2.40} &0.18  \\
{Reference}  &\textbf{-0.64}&0.13 & \textbf{2.50}& 0.18& &-0.43&0.13 &\textbf{2.39} &0.18  \\
\hline
\end{tabular*}\end{table*}

The mean age of the $D2_{test}$ participants was $71.8 \pm 5.4$ years (range: $51.7 - 97.0$ years). The number of female participants was $586$ ($53.1$ \%). Tract-specific average volumes and diffusion measures are provided in Supplementary Table~\ref{Stab1}. 

\begin{figure*}[!ht]
  \centering
   \includegraphics[scale=0.58]{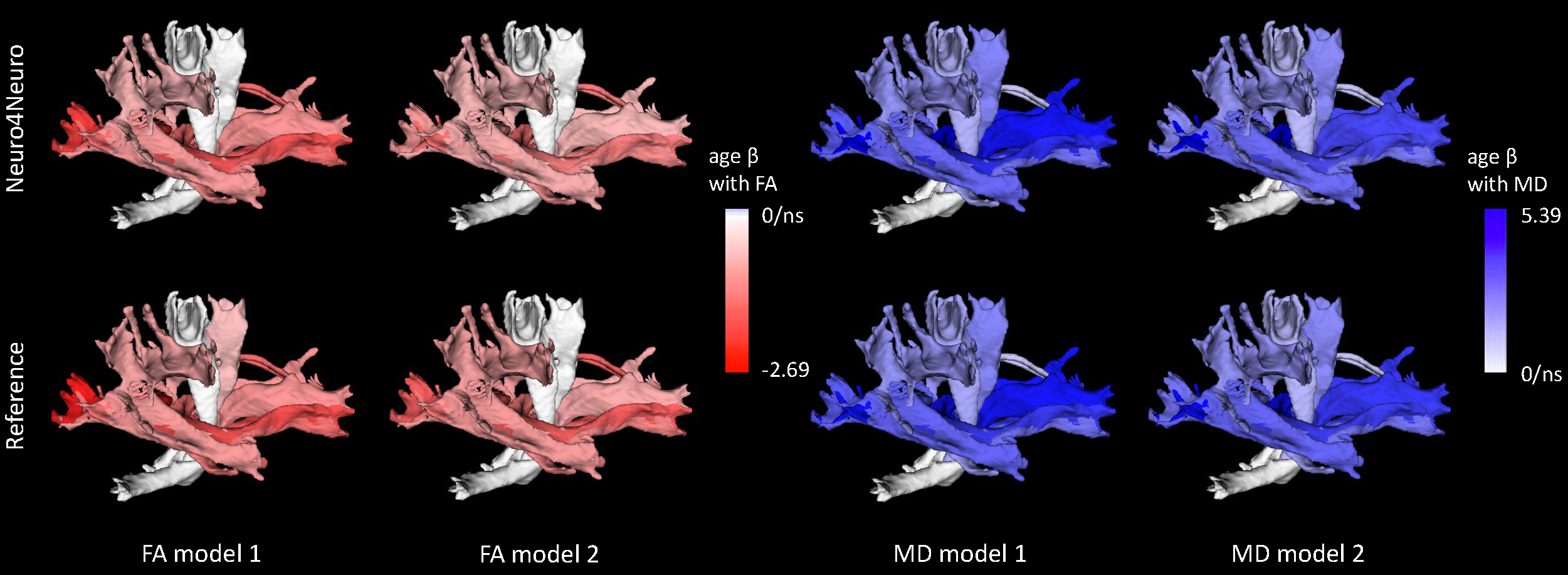}
\caption{Regression coefficients ($\beta$) for the associations of age with tract-specific fractional anisotropy (FA) and mean diffusivity (MD) determined by the proposed method (Neuro4Neuro) and the reference method. The shown tract segmentation was generated by Neuro4Neuro for a female participant (age = 79 years). Models were adjusted for sex, intracranial volume, and tract-specific volume (Model 2). Non-significant (ns) associations are shown in white.}\label{fig7}
\end{figure*}

The associations between age and tract-specific diffusion metrics obtained with the proposed method corresponded to those obtained with the reference method (FA, MD: Table~\ref{tab2} and Figure~\ref{fig7};  L1, RD and MO: Supplementary Table~\ref{Append_tab2}). In all models, significant degradation of the microstructural organization with aging (i.e., a decrease in FA and an increase of MD) was observed in the association tracts, commissural tracts and limbic tracts. For the sensorimotor tracts, which are known to be relatively spared from age-related deterioration \citep{de2015tract}, we found only weak correlations between age and FA or MD and relatively high associations between age and the mode of anisotropy (MO). Although in the STR tract both methods found a similar regression coefficient for the association between age and FA, this association was the only one that was significant for the reference method but not for the proposed method. Adjusting for tract volume in model 2 resulted in a slight attenuation in the associations of most tracts (except for the limbic tracts), which indicates that the loss of microstructure could be partially explained by tract atrophy. This effect was relatively larger for the proposed method than for the reference method, i.e., larger absolute changes in regression coefficient ($\beta$).

\subsubsection{Generalization to a dementia dataset}
\label{sssec:3.3.4.1}
\begin{figure*}[!ht]
  \centering
   \includegraphics[scale=0.48]{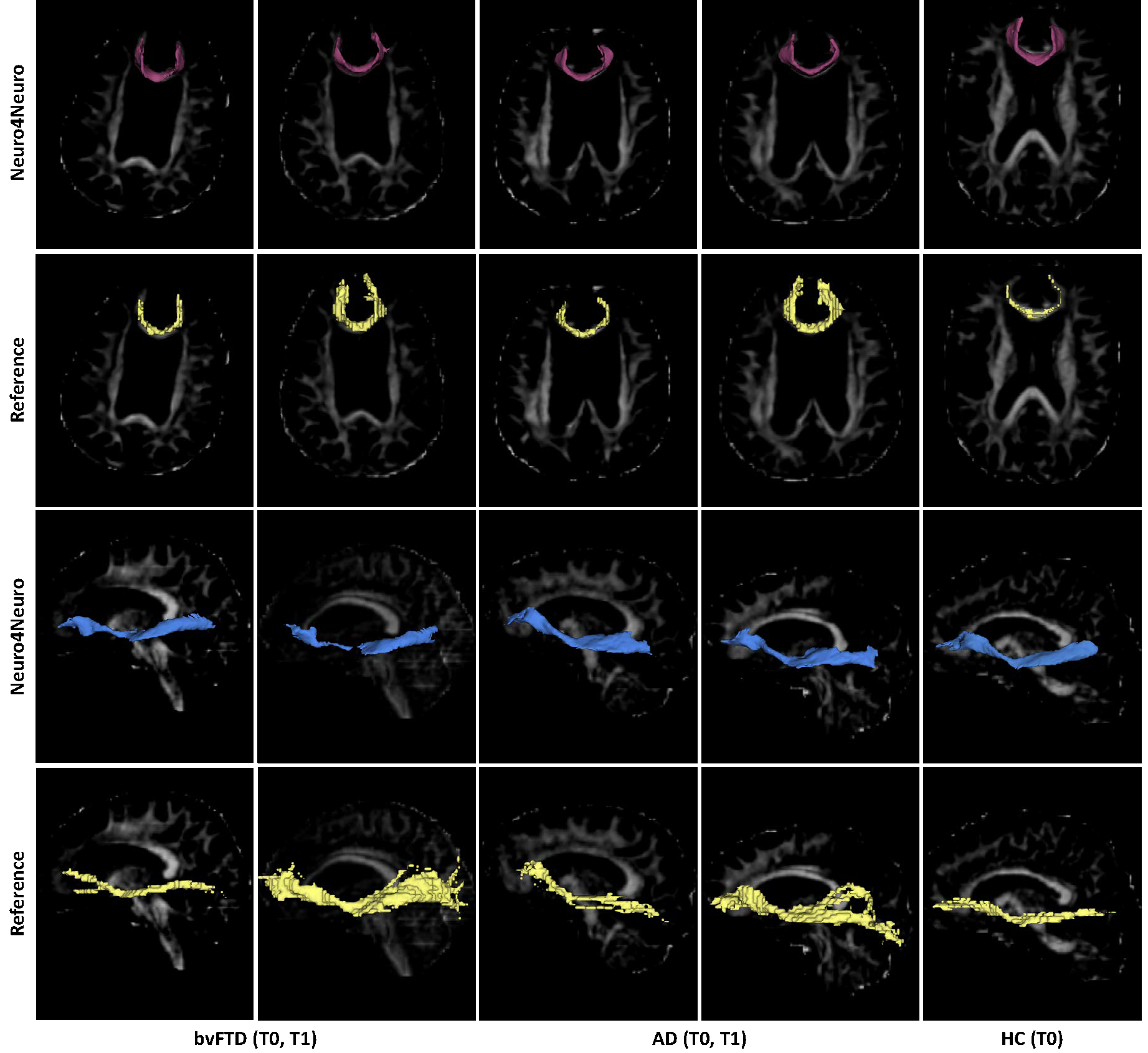}
    \caption{Tract segmentations by the proposed method on a dementia study dataset ($D4$) for the forceps minor (FMI, pink) and the inferior fronto-occipital fasciculus (IFO, blue). Reference method results are shown in yellow. Three participants with representative performance were selected: a patient with behavioral-variant frontotemporal dementia (bvFTD), a patient with Alzheimer's disease (AD) and a healthy control (HC). Scans were obtained at study baseline (T0) and one-year follow-up (T1).}\label{fig8}
\end{figure*}

The proposed method yielded visually good tract segmentations for controls and patients with bvFTD or AD; Figure \ref{fig8} shows examples of 3D tract-volume renderings overlaid on the corresponding FA images. We selected the FMI and IFO tracts for visualization as they are known to be involved in dementia \citep{rascovsky2011sensitivity}, belong in distinct tract categories and had different segmentation complexity, i.e., the thin and arch-shaped FMI tract is more difficult to segment than the long and straight IFO tract. We observed that the proposed method generally segmented the entire tracts accurately. The tracts were of consistent shape and size across participant groups and across time-points. The tracts shapes of the proposed method were generally similar to those of the reference method, although those of the latter tended to have a larger structure at follow-up and more often included parts of other tracts.     

\subsubsection{Groupwise differentiation of bvFTD and AD}
\label{sssec:3.3.4.2}

\begin{figure*}[!ht]
  \centering
     \subfigure[]{
  	 \includegraphics[scale=0.49]{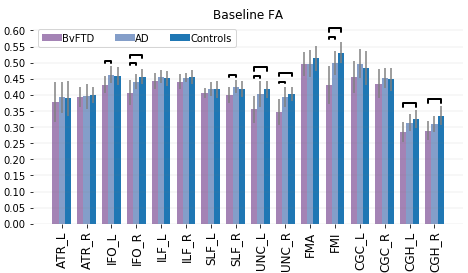}}
  	 \quad
     \subfigure[]{
  	 \includegraphics[scale=0.49]{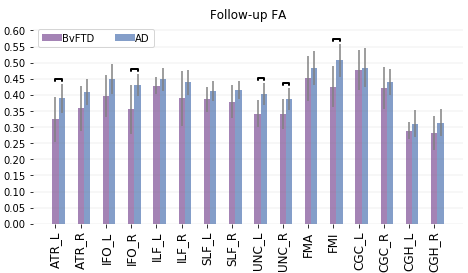}}
     
     \subfigure[]{
  	 \includegraphics[scale=0.49]{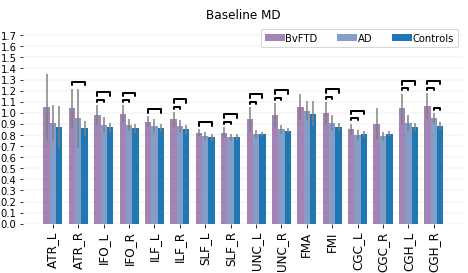}}
  	 \quad
     \subfigure[]{
  	 \includegraphics[scale=0.49]{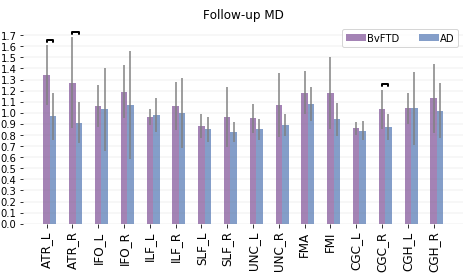}}
 \caption{WM microstructural abnormalities in behavioral variant frontotemporal dementia  (bvFTD) and Alzheimer's disease (AD) at baseline and follow-up. FA: fractional anisotropy, MD: mean diffusivity ($10^{-3} mm^{2}/s$). Error bars show standard deviations; bold brackets show significant difference between groups ($p<0.05$).}\label{fig9}
\end{figure*}

At baseline, microstructural differences between patient (bvFTD, AD) and control groups  were observed in several tracts (FA, MD: Figure \ref{fig9}; L1, RD: Figure \ref{Sup_fig1}). Tract-specific measurements were more abnormal in bvFTD than in AD in all tract categories, consistent with the results by \cite{meijboom2019exploring}. For bvFTD, most pronounced abnormalities were seen in the FMI, CGH, IFO and UNC tracts, while the FMA tract was the only tract in which WM microstructure was preserved. For AD, only CGH microstructure was found to be significantly different from controls. These findings are consistent with literature \citep{meijboom2019exploring, rascovsky2011sensitivity, laforce2013behavioral, mckhann2011diagnosis}.  

At follow-up, microstructural differences between bvFTD and AD groups were observed in fewer tracts than at baseline (FA, MD: Figure \ref{fig9}; L1, RD: Figure \ref{Sup_fig1}). Tract-specific measurements were more abnormal in bvFTD than in AD, which was significant for one or more metrics in the ATR, UNC, FMI, right IFO, and right CGC tracts. Tract abnormalities at follow-up were consistent with those at  baseline and also those of the study we replicated \citep{meijboom2019exploring}.

\section{Discussion}
\label{s:4}
We present a 3D-CNN-based method for direct WM tract segmentation: Neuro4Neuro. The method was developed and applied on a large set of dMRI images, yielding a high reproducibility and a good accuracy. We demonstrate that it was generalizable to a patient dataset acquired with different scanner hard- and software and a different MR imaging protocol. We assessed the applicability of the proposed WM tract segmentation method in preclinical and clinical research, by performing proof-of-principle experiments of WM microstructure degeneration in aging and WM microstructural differences between bvFTD and AD. Results of those analyses were found to be in line with those reported in literature.

The main strengths of our approach for WM tract segmentation are its performance and its applicability. First, measurements obtained with our method showed high accuracy, reproducibility and correlation with age and disease, as shown by extensive validation experiments using large and independent evaluation cohorts. Second, regarding applicability, our method both accelerates and simplifies WM tract segmentation. A tremendous acceleration is achieved as our method reduces the time required for tract segmentation from roughly 35 hours using tractography-based methods to only 0.5 seconds per tract per scan. A part of this speed-up could also be achieved by a GPU implementation of tractography-based methods \citep{hernandez2018using}. In addition, our method simplifies tract segmentation by using an end-to-end learning approach, which avoids many separate steps such as parcellation, atlas registration and fiber tracking. Also, the application of the method does not require any special hardware, but can be used on a normal workstation with a CPU or a GPU. The runtime of input preparation, i.e., diffusion tensor estimation, is 31 seconds per scan on a CPU node. The subsequent segmentation took 0.49 seconds from loading diffusion tensor images to save the segmented results. Since WM tract microstructure has shown to be valuable in several studies \citep{lovden2013dimensionality, white2009white, jones2005effect, smith2006tract}, it is essential to have a method for tract segmentation that can reliably characterize WM microstructure and is easy to apply. Therefore, as the proposed method meets both these criteria, we believe that it can be beneficial for both clinical practice (e.g., monitoring neurodegeneration in individuals for diagnosis or for a clinical trial) and large-scale population studies (e.g., studying neurodegeneration in aging).

Based on optimization experiments (Section \ref{ssec:2.2}), we propose an encoder-decoder CNN with skip connections that is optimized with the Adam algorithm based on a weighted inner product loss function. Segmentation accuracy was not improved by adding novel elements to the network architecture, i.e., convolutional re-samplings and residual functions. It should be noted that we did not control for the total number of parameters in these experiments as our purpose was to optimize accuracy. We adopted a 3D network architecture since compared to 2D CNN methods, 3D methods are expected to reduce the number of required training samples \citep{milletari2017hough}, to increase segmentation accuracy \citep{isensee2018nnu}, and to better exploit 3D spatial information in each estimation step which would not be achieved when 2D slices are processed independently. Regarding input, we found that using just the diffusion tensor image yielded optimal performance. Spatial information generally slightly increased segmentation accuracy, but its added value was only marginal. Model optimization was performed on data of one tract. Potentially the results would improve when optimizing on each tract specifically or all tracts combined. An alternating-update strategy was adopted for the optimization on all components rather than a full gradient descent, with the aim of exploring possible combinations to further improve from the current best configuration based on prior knowledge.

The proposed method yielded good segmentation performance in terms of accuracy, reproducibility and generalizability. The average segmentation accuracy was $DC=0.74$ over 25 tracts, with the best performance in the medial lemniscus tract ($DC=0.84$) (Figure~\ref{fig5}). Relative accuracy between individual tracts followed a similar trend as reported in the literature \citep{oishi2009atlas, jin2014automatic, wasserthal2018tractseg}. Segmentation of the ILF for example yielded lower accuracy than other tracts, which may due to the fact that up to five tracts pass through the temporal lobe ``bottleneck'' resulting in ambiguities in crossing-fiber analysis \citep{maier2017challenge}. Also, since the DC is very sensitive to the size of the object, a small and thin object will always have a lower DC value such as the FMI tract.

In addition, Neuro4Neuro achieved a high reproducibility both in terms of voxel-wise agreement of segmentations ($\kappa$) and correspondence between tract-specific measures ($\epsilon, R^2$) (Figure~\ref{fig6}). There are two main reasons that the proposed approach significantly improves reproducibility. First, shape and spatial priors, and ``free-form'' parameters for feature extraction and classification are globally optimized in an objective-driven manner on a large-scale dataset, which has been widely demonstrated to outperform manual-crafted features and predefined models. Second, the reference method  outputs less consistent spatial segmentations, whereas it provides reproducible diffusion measures. This is the main reason that the improved reproducibility is more remarkable for tract volume and spatial correspondence ($\kappa$) while smaller for diffusion measures. Apparently, diffusion measures are more robust to variations in the segmentation. In addition, the reproducibility of Neuro4Neuro is also similar to those reported for a longitudinal method by \cite{yendiki2016joint} and higher than those reported for manual segmentations by \cite{kaur2014reliability,wakana2007reproducibility}. High reproducibility is especially important for analysis of longitudinal data and for studies across different groups or datasets. In general, a method with high reproducibility requires a smaller sample size or less time-points to achieve the same statistical power \citep{yendiki2016joint}. Hence, we argue that the proposed method is a reliable tool for analysis of WM microstructure.

The generalizability of the proposed method was demonstrated by an evaluation on an external patient dataset (Iris dataset; Figure~\ref{fig8}, Figure~\ref{fig9}). Overall, our method generalized very well to this dataset, showing good segmentations for most tracts, as well as consistent tract architectures across participant groups and time-points. Although this test dataset was completely different from the optimizing data regarding patient populations, MRI scanners, scanning protocols and tensor estimation algorithms, only a subtle deterioration of segmentation performance was noticed. First, we saw a slight increase in the number of false positive points mainly at skull-voxels of the FMI segmentation. Second, we noticed that only for the IFO tract, the structure was occasionally disconnected at the thin and ``twisted'' middle section. We suspect that this was mainly due to the brain tissue mask that was applied to the training data as a preprocessing step but not to the Iris data. This tissue mask was obtained with a segmentation method that was specially optimized for the Rotterdam Study data \citep{vrooman2007multi} and therefore could not be obtained for the Iris data. Also, given the observation that our learning-based method showed better generalization than the non-learning based reference method, it can be considered unnecessary to retrain the model for this different data distribution, which is another advantage for future applications.

In two proof-of-principle experiments, we demonstrated the applicability of our method in WM microstructure analysis for epidemiological and clinical studies. The first experiment showed a widespread reduction of microstructural organization with aging (Table \ref{tab2}), which was consistent with previously published results \citep{de2015tract}. Adjusting for tract volume resulted in attenuated associations for the proposed method to a larger extent than those for the reference method. This means that for the proposed method tract volume has an increased confounding in the associations between age and tract-specific diffusion measures, which is probably due to the more robust volume measurements that have a higher correlation with age and diffusion measures and also allows the investigation of WM macrostructure. The second experiment showed the method's performance in differentiation of different diseases underlying dementia (i.e., AD and bvFTD) based on tract-specific WM microstructure measurements (Figure~\ref{fig9}, Figure~\ref{Sup_fig1}). We found that diffusion measurements in all tract categories were more abnormal in bvFTD than in AD. Since for both the normal and the dementia population, the found associations were in line with those reported in the literature \citep{de2015tract,meijboom2019exploring, rascovsky2011sensitivity, laforce2013behavioral, mckhann2011diagnosis}, we believe that the proposed method can be applied to such epidemiological and clinical studies as well. The method is designed for analysis of diffusion measures over entire tracts and uses a voxel-wise classification strategy. Therefore, it does not lend itself for along-the-tract analyses, which would allow for detecting local effects that may be lost during averaging.

A challenge in WM tract segmentation is that there is no ``gold standard'' for tract in vivo \citep{crick1993backwardness}. Therefore, we quantified segmentation accuracy with respect to a reference standard \citep{de2015tract}. It is non-trivial to obtain a reliable reference standard because of high inter-subject variability in tract anatomy and a lack of consensus in tract definitions \citep{sydnor2018comparison}. Also, because methods are often optimized for a specific use-case, it is challenging to compare performances. 

The reference standard in this work was based on probabilistic tractography and thresholding using a reproducibility-based metric \citep{de2015tract}. As training labels, this approach is limited by some incomplete and disconnected segmentations. In addition, we observed a relatively high intra-subject agreement for the reference segmentation in the central brain regions but this tended to diverge more towards cortical regions for some tracts (Figure~\ref{fig8}). We suspect this is inherent to a method that does not enforce shape consistency. Deterministic tractography based segmentation approaches might have served as an alternative reference standard. Deterministic tractography methods generally have a higher fiber validity, while their lower scores on volume-orientated metrics could introduce other kind of variations during training for instance in some offshoots of tracts \citep{maier2017challenge,poulin2019tractography}. Despite these limitations in the reference segmentations, we expect that they did not have much effect on the performance of our method. Although segmentation accuracy (DC) with this reference standard could be slightly lower than values reported in other articles computed with a more smooth reference standard, we demonstrated that our method can segment complete tracts and has high intra- and inter-subject consistency.

Despite recent advances in higher-order diffusion models \citep{hyde2019white}, we use the relatively simple DTI model. The major advantage of using the DTI model is that it enables our method to be applicable to clinical data, which like the datasets in this article usually do not support more than two fiber populations \citep{behrens2007probabilistic}. Using peaks of the fiber orientation distribution function (fODF) as input gave significantly inferior results on our dataset. We preferred using the diffusion tensor image over using the raw diffusion-weighted MRI data, as this is more efficient in memory and computation time. In addition, this enables combination of different datasets since the dimensionality of the diffusion tensor does not depend on the number of slices or diffusion weighted gradients.

We performed a pilot experiment on our optimization dataset to compare Neuro4Neuro with an existing CNN-based WM tract segmentation method \citep{wasserthal2018tractseg}. Results are however not included in this manuscript, since we failed to replicate the performance reported by the literature on our dataset and since optimizing the approach beyond trying the default implementation exceeded the scope of the current work. For this, a common evaluation framework as for instance provided by challenges would be beneficial.

We demonstrated the generalizability of our method on a dementia dataset, but it would also be interesting to evaluate the performance on other diseases. We expect that our method has good generalizability to data of patients with neurodegenerative diseases, e.g., Parkinson's disease and Huntington's disease. However, for disease with large and abrupt changes in brain diffusion such as brain tumors, further refinement of the method is probably required, which would be an interesting future research area. In the experiments we found that Neuro4Neuro had good generalizability to different preprocessing pipelines and MRI acquisition protocols, e.g., when training on 1.5T MRI and testing on 3T MRI. Generalizability to other datasets with other b-values and number of directions has yet to be performed. 

We conclude that the proposed WM tract segmentation method, Neuro4Neuro, improves reproducibility compared to the reference method, and provides a reliable generalizable method for analyzing WM microstructure. In addition, the proposed method is orders of magnitude faster. To our best knowledge this is the first deep learning based method for WM tract segmentation that is developed and evaluated on such a large-scale dataset. Our method can lead toward a faster, more lightweight way of WM tract segmentation and WM microstructure analysis.

\section*{Acknowledgements}
The authors are grateful to SURFsara for the processing time on the Dutch national supercomputer (\url{www.surfsara.nl/systems/cartesius}). 

B. Li and W.J. Niessen are supported by Medical Delta Diagnostics 3.0: Dementia and Stroke. M. de Groot acknowledges funding from the EU Horizon 2020 project EuroPOND (666992). E.E. Bron acknowledges support from the Netherlands CardioVascular Research Initiative (Heart-Brain Connection: CVON2012-06, CVON2018-28) and the Dutch Heart Foundation (PPP Allowance, 2018B011). In addition, M. de Groot has a financial interest in the GSK company. The GSK had no role in this study.


\section*{Reference}

\bibliographystyle{apa}
\DeclareRobustCommand{\de}[3]{#3}
\bibliography{WM_tract_segmentation_N4N.bib}

\appendix
\section*{Appendix A. Supplementary material}
\beginsupplement

\begin{figure*}[!ht]
  \centering
   \includegraphics[scale=0.46]{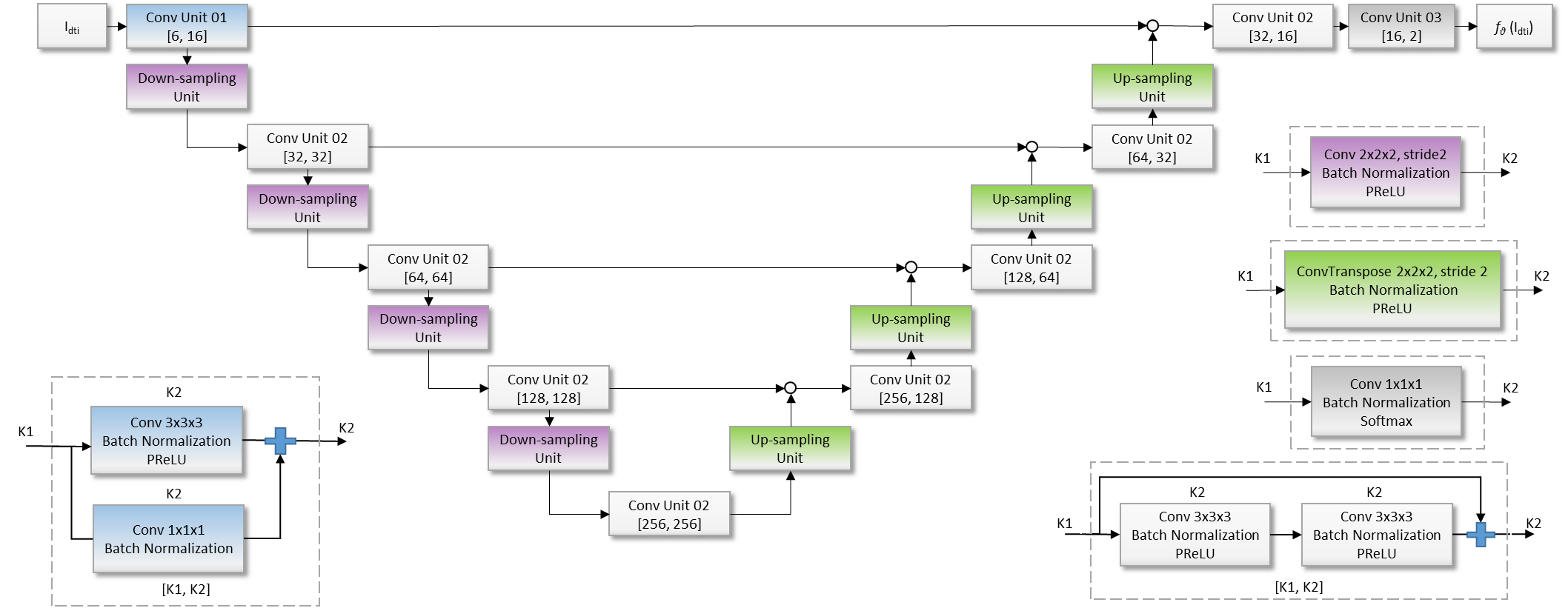}
  \caption{The {\itshape Ext-architecture}. The colored boxes in the lower left and right corners provide a legend for corresponding units in the network architecture, where $[k_1, k_2]$ are the number of convolution kernels in those layers. Abbreviations: Conv = convolution, PRelu = parametric rectified linear units. The circles in the decoder path indicate concatenation operations. The plus in the legends indicates residual addition.}\label{Sfig1}
\end{figure*}

\begin{table*}[!ht]
\scriptsize
\centering
\caption{Average tract-specific volume and diffusion measurements determined by Neuro4Neuro. Volume in ml; mean diffusivity (MD), axial diffusivity (L1) and radial diffusivity (RD) $\times 10^{-3} mm^2/s$. Tract abbreviations refer to the Figure~\ref{fig5}. SD, standard deviation. }\label{Stab1}
\begin{tabular*}{0.99\textwidth}
{@{}@{\extracolsep{\fill}}cccccccccccccccccc@{}}
\hline
\multicolumn{1}{c}{ }  &
\multicolumn{2}{c}{Volume} &
\multicolumn{1}{c}{}   &
\multicolumn{2}{c}{FA} &
\multicolumn{1}{c}{}   &
\multicolumn{2}{c}{MD} &
\multicolumn{1}{c}{}   &
\multicolumn{2}{c}{L1} &
\multicolumn{1}{c}{}   &
\multicolumn{2}{c}{RD} &
\multicolumn{1}{c}{}   &
\multicolumn{2}{c}{MO} \\
\cline{2-3}\cline{5-6}\cline{8-9}\cline{11-12}\cline{14-15}\cline{17-18}
 Tract &Mean &SD & &Mean &SD &&Mean &SD &&Mean &SD & &Mean &SD &&Mean &SD\\
 \hline
{\itshape Association} \\
{ATR}   & 6.37& 0.90&&  0.36&0.02&&  0.79&0.05&& 1.09&0.06&& 0.63&0.05&&  0.41&0.05\\
{IFO}   &5.47&0.54&& 0.42&0.02&& 0.82&0.05&& 1.22&0.05&& 0.62&0.05&& 0.52&0.06 \\
{ILF}   &5.91&0.52&&  0.42&0.02&& 0.79&0.04&& 1.18&0.04&& 0.60&0.04&& 0.50& 0.05\\
{PTR}   &3.94&0.42&& 0.39&0.02&& 0.84&0.06&& 1.21&0.07&& 0.65&0.06&& 0.41&0.07\\
{SLF}   &9.50&1.19&& 0.38&0.02&& 0.73&0.04&& 1.04&0.04 &&0.58 &0.04 &&0.30 &0.06\\
{UNC}   &2.41&0.36&& 0.38&0.02&& 0.77&0.03&& 1.11&0.04&& 0.60& 0.03&& 0.57&0.05\\
{\itshape Commissural} \\
{FMA}   &7.09&0.83&&  0.51&0.03&& 0.83&0.05&& 1.36&0.06&& 0.57&0.05&& 0.68& 0.06\\
{FMI}   &4.08&0.60&& 0.47&0.03&& 0.80&0.04&& 1.28&0.04&& 0.57&0.04&& 0.57&0.07  \\
{\itshape Limbic} \\
{CGC}   &1.22&0.15&&  0.43&0.04&& 0.73&0.02&& 1.10&0.04&& 0.54&0.03&& 0.54&0.11\\
{CGH}   &0.97&0.12&&  0.32&0.03&& 0.78&0.03&& 1.05&0.04&& 0.64&0.04&& 0.51& 0.08 \\
{\itshape Sensorimotor} \\
{CST}  &7.22&0.89&&  0.47&0.02&& 0.72&0.03&& 1.11&0.04&& 0.52&0.03&& 0.54&0.05 \\
{MCP}  &4.35&0.73&&    \\
{ML}   &2.19&0.22&&  0.44&0.02&&  0.77&0.02&& 1.16&0.03&& 0.58&0.02&& 0.54&0.05\\
{STR}  &6.03&0.70&&  0.42& 0.02&& 0.71& 0.03&& 1.06& 0.05&& 0.54& 0.03&& 0.43& 0.08 \\
\hline
\end{tabular*}
\end{table*}

\begin{table*}[!ht]
\scriptsize
\centering
\caption{Associations between age and tract-specific diffusion measures. Tract abbreviations refer to the Figure~\ref{fig5}. Values ($\times 10^{-3}$) represent regression coefficients ($\beta$) and their standard error ($std.error$) for change in axial diffusivity (L1), radial diffusivity (RD) or mode of anisotropy (MO) per year increase in age, adjusted for sex and ICV (and additionally for tract-specific WM volume in Model 2). Significant associations at Bonferroni corrected threshold $P-value=6.0\times10^{-4}$ are shown in \textbf{bold}. }\label{Append_tab2}
\begin{tabular*}{0.99\textwidth}
{@{}@{\extracolsep{\fill}}cccccccccccccc@{}}
\hline
\multicolumn{1}{c}{ }  &
\multicolumn{6}{c}{Model 1} &
\multicolumn{1}{c}{}   &
\multicolumn{6}{c}{\makecell{Model 2: \\Model 1 + WM volume}} \\
\cline{2-7}\cline{9-14}
 Tract &L1 $\beta$ &L1 $std.error$ &RD $\beta$ &RD $std.error$&MO $\beta$& MO $std.error$& &L1 $\beta$ &L1 $std.error$&RD $\beta$ &RD $std.error$ &MO $\beta$ &MO $std.error$ \\
 \hline
{\itshape Association} \\
{ATR}    \\
{Neuro4Neuro}  &\textbf{5.34}& 0.27&\textbf{4.67}& 0.24& 0.35 &0.26& & \textbf{4.35}& 0.27&\textbf{3.57}& 0.25&\textbf{1.11}&0.28 \\
{Reference}  &\textbf{5.29}& 0.26&\textbf{4.68}&  0.24& 0.22 &0.26& & \textbf{4.66}& 0.26&\textbf{3.92}& 0.24&0.87&0.26 \\

{IFO}    \\
{Neuro4Neuro}  &\textbf{4.12}& 0.23&\textbf{4.51}& 0.23& \textbf{-3.29} &0.30& & \textbf{3.69}& 0.25&\textbf{3.71}& 0.24&\textbf{-2.46}&0.32\\
{Reference}   &\textbf{3.64}& 0.24&\textbf{4.33}& 0.23& \textbf{-3.57} &0.31& & \textbf{3.67}& 0.24&\textbf{4.15}& 0.23&\textbf{-3.19}&0.30 \\

{ILF}    \\
{Neuro4Neuro}  &\textbf{3.36}& 0.21&\textbf{2.95}& 0.21& 0.19 &0.28& & \textbf{3.36}& 0.21&\textbf{2.85}& 0.21&0.15&0.29 \\
{Reference}    &\textbf{3.43}& 0.23&\textbf{3.05}& 0.22& 0.25 &0.28& & \textbf{3.43}& 0.23&\textbf{3.05}& 0.22&0.25&0.28 \\

{PTR}    \\
{Neuro4Neuro}  &\textbf{5.55}& 0.32&\textbf{5.23}& 0.28& \textbf{-2.34} &0.36& & \textbf{5.65}& 0.32&\textbf{5.26}& 0.29&\textbf{-2.51}&0.37 \\
{Reference}    &\textbf{5.21}& 0.37&\textbf{4.92}& 0.31& \textbf{-1.87} &0.37& & \textbf{5.37}& 0.36&\textbf{4.95}& 0.31&\textbf{-1.78}&0.37 \\

{SLF}    \\
{Neuro4Neuro}  &\textbf{2.18}& 0.18&\textbf{2.12}& 0.21& \textbf{-1.11} &0.31& & \textbf{2.19}& 0.18&\textbf{2.07}& 0.21&-1.01&0.30 \\
{Reference}    &\textbf{2.00}& 0.18&\textbf{2.19}& 0.21& \textbf{-1.31} &0.32& & \textbf{2.01}& 0.18&\textbf{2.25}& 0.20&\textbf{-1.41}&0.30\\

{UNC}    \\
{Neuro4Neuro}  &\textbf{2.38}& 0.19&\textbf{3.05}& 0.17& \textbf{-1.79} &0.28& & \textbf{2.49}& 0.19&\textbf{2.80}& 0.17&\textbf{-1.14}&0.26 \\
{Reference}    &\textbf{2.38}& 0.17&\textbf{3.08}& 0.18& \textbf{-2.02} &0.28& & \textbf{2.53}& 0.17&\textbf{2.91}& 0.17&\textbf{-1.51}&0.25 \\

{\itshape Commissural} \\
{FMA}    \\
{Neuro4Neuro}  &\textbf{2.45}& 0.32&\textbf{3.69}& 0.29&\textbf{-2.17} &0.32& & \textbf{2.35}& 0.33&\textbf{2.34}& 0.27&-0.39&0.29 \\
{Reference}    &\textbf{2.05}& 0.32&\textbf{4.01}& 0.31&\textbf{-2.54} &0.34& & \textbf{2.12}& 0.33&\textbf{3.20}& 0.30&\textbf{-1.47}&0.33 \\

{FMI}     \\
{Neuro4Neuro}  &\textbf{1.35}& 0.22&\textbf{3.72}& 0.23&\textbf{-4.59} &0.38& & \textbf{1.84}& 0.24&\textbf{2.29}& 0.22&\textbf{-2.11}&0.36 \\
{Reference}   &\textbf{1.10}& 0.24&\textbf{3.70}& 0.24&\textbf{-4.75} &0.39& & \textbf{1.66}& 0.23&\textbf{2.98}& 0.23&\textbf{-3.30}&0.37 \\

{\itshape Limbic} \\
{CGC}    \\
{Neuro4Neuro}  & -0.35&0.22 &\textbf{1.81}&0.17&\textbf{-6.79}&0.55 &&-0.24&0.22&\textbf{1.71}&0.17 &\textbf{-6.48}&0.53 \\
{Reference}   & -0.55&0.21 &\textbf{1.89}&0.19&\textbf{-8.06}&0.57 &&-0.62&0.21&1.94&0.18 &\textbf{-8.25}&0.56\\

{CGH}     \\
{Neuro4Neuro}  & \textbf{1.18}&0.21 &\textbf{2.20}&0.19&-0.86&0.44 &&\textbf{1.22}&0.22&\textbf{2.12}&0.19 &-0.41&0.43 \\
{Reference}    &  \textbf{1.13}&0.24 &\textbf{2.36}&0.24&-1.02&0.44 &&\textbf{1.14}&0.24&\textbf{2.35}&0.24 &-0.95&0.41 \\

{\itshape Sensorimotor} \\
{CST}    \\
{Neuro4Neuro}  & \textbf{2.60}&0.23 &\textbf{1.09}&0.15&\textbf{1.07}&0.30 &&\textbf{2.21}&0.24&\textbf{0.98}&0.16 &\textbf{1.16}&0.32 \\
{Reference}    & \textbf{2.46}&0.19 &\textbf{1.43}&0.15&0.66&0.31 &&\textbf{2.23}&0.19&\textbf{1.15}&0.15 &0.71&0.32\\

{MCP}     \\
{Neuro4Neuro}  & 1.35&0.45 &0.27&0.41&0.98&0.40    &&0.65&0.39&-0.26&0.38 &0.85&0.40\\
{Reference}    & 1.45&0.55 &0.05&0.49&1.24&0.45    &&0.68&0.48&-0.60&0.43 &1.15&0.45 \\

{ML}     \\
{Neuro4Neuro}  & 0.42&0.18 &0.05&0.12&\textbf{-1.56}&0.29    &&0.53&0.18&-0.01&0.12 &\textbf{-1.20}&0.28 \\
{Reference}    & 0.37&0.20 &0.22&0.13&\textbf{-1.59}&0.30    &&0.47&0.20&0.19&0.13 &\textbf{-1.35}&0.29 \\

{STR}     \\
{Neuro4Neuro}  &\textbf{3.27}& 0.23&\textbf{2.06}& 0.18&\textbf{2.96} &0.44& & \textbf{3.45}& 0.23&\textbf{1.88}& 0.19&\textbf{3.75}&0.44 \\
{Reference}    &\textbf{3.09}& 0.22&\textbf{2.21}& 0.19&\textbf{2.23} &0.41& & \textbf{3.17}& 0.22&\textbf{2.01}& 0.19&\textbf{2.63}&0.41 \\
\hline
\end{tabular*}
\end{table*}

\begin{figure*}[!ht]
  \centering
     \subfigure{
  	 \includegraphics[scale=0.49]{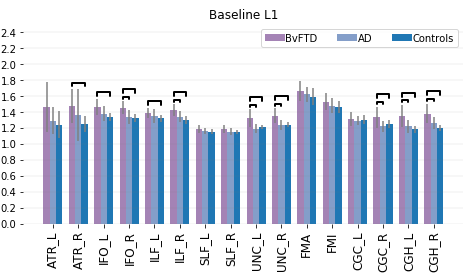}}
  	 \quad
     \subfigure{
  	 \includegraphics[scale=0.49]{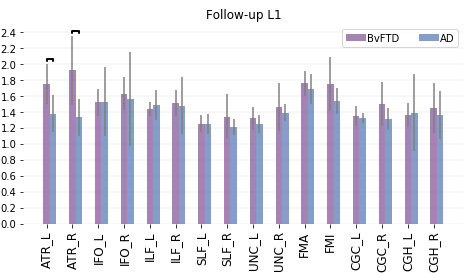}}
     
     \subfigure{
  	 \includegraphics[scale=0.49]{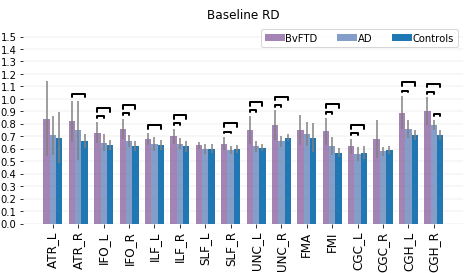}}
  	 \quad
     \subfigure{
  	 \includegraphics[scale=0.49]{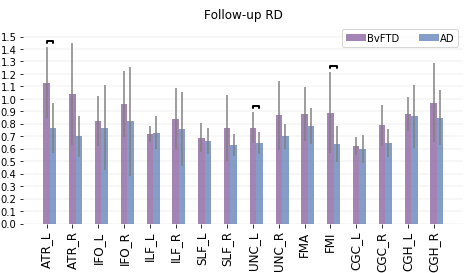}}
 	 \caption{WM microstructural abnormalities in bvFTD and AD. BvFTD, behavioural variant frontotemporal dementia. AD, Alzheimer's disease.  L1, axial diffusivity, $10^{-3} mm^{2}/s$. RD, radial diffusivity, $10^{-3} mm^{2}/s$. Error bars, standard deviations. The bold brackets indicate significant difference in groups ($p<0.05$).}\label{Sup_fig1}
\end{figure*}

\end{document}